\documentclass[preprint,aps,showpacs,nofootinbib,preprintnumbers,amsmath,amssymb,superscriptaddress]{revtex4-1}
\pdfoutput=1
\usepackage{amssymb}
\usepackage{epsfig}
\usepackage{latexsym}
\usepackage{graphicx}% Include figure files
\usepackage{subfigure}
\usepackage{dcolumn}% Align table columns on decimal point
\usepackage{bm}% bold math
\usepackage[usenames ,dvipsnames]{xcolor}
\usepackage[utf8]{inputenc}
\usepackage{slashed}
\usepackage[normalem]{ulem}
\usepackage{cleveref}
\definecolor{bostonuniversityred}{rgb}{0.8, 0.0, 0.0}

\begin{document}

\title{Anomalies in $B$ Decays and Muon $g-2$ from Dark Loops}

\author{
  Da~Huang\footnote{dahuang@ua.pt}
}
 \affiliation{Departamento de F\'{\i}sica da Universidade de Aveiro and CIDMA, \\
 Campus de Santiago, 3810-183 Aveiro, Portugal.}
\affiliation{National Astronomical Observatories, Chinese Academy of Sciences, Beijing, 100012, China}

\author{
  Ant\'{o}nio~P.~Morais\footnote{aapmorais@ua.pt}
}
 \affiliation{Departamento de F\'{\i}sica da Universidade de Aveiro and CIDMA, \\
 Campus de Santiago, 3810-183 Aveiro, Portugal.}

\author{Rui~Santos\footnote{rasantos@fc.ul.pt}}
\affiliation{Centro de F\'{\i}sica Te\'{o}rica e Computacional,
   Faculdade de Ci\^{e}ncias,\\
   Universidade de Lisboa, Campo Grande, Edif\'{\i}cio C8
  1749-016 Lisboa, Portugal.}
\affiliation{ISEL - Instituto Superior de Engenharia de Lisboa,\\
  Instituto Polit\'ecnico de Lisboa
 1959-007 Lisboa, Portugal.}

\date{\today}
\begin{abstract}
We explore a class of models which can provide a common origin for the recently observed evidence for lepton flavor universality violation in $b \to s l^+ l^- $ decays, the dark matter (DM) problem and the long-standing muon $(g-2)$ anomaly. In particular, both anomalies in the $B$ meson decays and the muon $(g-2)$ can be explained by the additional one-loop diagrams with DM candidates. We first classify several simple models according to the new fields' quantum numbers. We then focus on a specific promising model and perform a detailed study of both DM and flavor physics. %As a result, the dark matter candidate can be the real or imaginary component of a colorless complex scalar. 
A random scan over the relevant parameter space reveals that there is indeed a large parameter space which can explain the three new physics phenomena simultaneously, while satisfying all other flavor and DM constraints. % shows that the DM particle should be rather light, and the Yukawa couplings should be sizeable. 
Finally, we discuss some of the possible new physics signatures at the Large Hadron Collider.

\end{abstract}

%\pacs{95.35.+d, 13.85.Tp, 14.80.-j, 98.70.Sa,}
%\keywords{Dark Matter, Cosmic Rays, AMS-02 Experiment}
\maketitle

%%%%%%%%%%%%%%%%%%%%%%%%%%%%%%%%%%%%%%%%%%%%%%%%%%%%%%%%%%%%%%%%%%%%%%%%%%%%%%%%%%%%%%%%%%%%%%%%%%%%%%%%%%%%%%%%%%%%%%%%%%%%%%%%%%%%%%%%%%%%%%%%%%%%%%%%%%%%%%%%%%%%%%%%%%
\section{Introduction}
\label{s1}
%People are always pursuing the New Physics (NP) beyond the Standard Model (SM). 
One of the most recent hints of New Physics (NP) comes from the observed anomalies in the semileptonic decay rates of the $B$ meson, which suggests a violation of lepton flavor universality. 
Concretely, the most precise measurement of the ratios of the exclusive branching fractions, $R(K^{(*)}) = {\cal B}(B\to K^{(*)}\mu^+\mu^-)/{\cal B}(B \to K^{(*)}e^+ e^-)$, is the one by the LHCb Collaboration~\cite{Aaij:2019wad,Aaij:2017vbb}, with the following values
\begin{eqnarray}
R(K)  = 0.846^{+0.060+0.016}_{-0.054-0.014} \,,\quad\quad q^2 \in [1.1,6] {\rm GeV}^2\,,
\end{eqnarray}
and
\begin{eqnarray}
R(K^*) = \left\{ \begin{array}{cc}
0.660^{+0.110}_{-0.070} \pm 0.024\,, & q^2 \in [0.045,1.1] {\rm GeV}^2\,,\\
0.685^{+0.113}_{-0.069} \pm 0.047\,, & q^2 \in [1.1,6] {\rm GeV}^2\,.
\end{array}\right.
\end{eqnarray}
where $q^2$ is the dilepton mass squared in the processes, while the corresponding  Standard Model (SM) predictions are~\cite{Hiller:2003js,Bordone:2016gaq}
\begin{eqnarray}
R(K) = 1.0004(8)\,, \quad\quad q^2 \in [1.1,6] {\rm GeV}^2\,,
\end{eqnarray}
and
\begin{eqnarray}
R(K^*) =\left\{ \begin{array}{cc}
0.920\pm 0.007\,, & q^2 \in [0.045,1.1] {\rm GeV}^2\,, \\
0.996 \pm 0.002\,, & q^2 \in [1.1,6] {\rm GeV}^2\,.
\end{array}
\right.
\end{eqnarray}
More recently, the Belle Collaboration has published their measurements on these two important quantities in Refs.~\cite{Abdesselam:2019wac,Abdesselam:2019lab},  with larger error bars compared with the LHCb results.
One should note that the quantities $R(K^{(*)})$ are very clean probes of NP because the theoretical and experimental uncertainties related to the hadronic matrix elements cancel out~\cite{Hiller:2003js}.
Further evidence supporting this $B$ physics anomaly has been obtained by measuring other observables in rare $B$ meson decays, such as the differential branching ratios~\cite{Aaij:2014pli, Aaij:2015esa, Wei:2009zv} and angular distribution observables~\cite{Aaltonen:2011ja, Khachatryan:2015isa, Abdesselam:2016llu, Lees:2015ymt, Aaij:2015oid, Wehle:2016yoi, Sirunyan:2017dhj, Aaboud:2018krd} in the processes $B\to \phi \mu^+\mu^-$ and $B\to K^{(*)} \mu^+ \mu^-$, which have also shown deviations from their SM predictions. Note that all anomalies are associated with the transition $b\to s\mu^+\mu^-$. In order to reconcile these discrepancies, many models have been proposed. One such type of models have lepton universality violation at tree level by introducing a $Z^\prime$~\cite{Buras:2013qja,Gauld:2013qja, Altmannshofer:2019xda, Lebbal:2020sqb, Capdevila:2020rrl} or a leptoquark~\cite{Bauer:2015knc, Angelescu:2018tyl, Angelescu:2019eoh, Balaji:2019kwe, Crivellin:2019dwb, Saad:2020ucl, Fuentes-Martin:2020bnh}, see {\it e.g.}, Ref.~\cite{Capdevila:2017bsm} for a review and references therein. One can also interpret the experimental data by one-loop penguin and box diagrams involving new exotic particles~\cite{Gripaios:2015gra, Arnan:2016cpy, Arnan:2019uhr, Hu:2019ahp, Hu:2020yvs}.

Besides the above NP signals in $B$ meson decays, there are other important hints like the long-standing low-energy flavor anomaly involving the measurement of the anomalous magnetic moment of the muon, $(g-2)_\mu$~\cite{PDG,Gorringe:2015cma}. The most recent prediction of this quantity in the SM~\cite{Blum:2018mom} has shown a $3.7\sigma$ discrepancy from the experimental measurement~\cite{Bennett:2006fi}:
\begin{eqnarray}\label{g2Val}
\Delta a_\mu = a_\mu^{\rm exp} - a_\mu^{\rm SM} \simeq (27.4\pm 7.3)\times 10^{-10}\,,
\end{eqnarray}
where the error is obtained by combining the theoretical and experimental uncertainties.
In the near future, a great reduction in the experimental uncertainty is expected, with the results from the experiments at J-PARC~\cite{J-PARC} and Fermilab~\cite{Grange:2015fou}. A further demand for NP arises from the increasing number of experiments pointing to the existence of dark matter (DM) in our Universe~\cite{Bergstrom:2012fi, PDG, Ade:2015xua}. However, despite the great experimental and theoretical efforts in detecting DM particles~\cite{Bergstrom:2012fi} during the last decades, the nature of DM remains a mystery in particle physics. The DM problem has already been investigated in various models~\cite{Vicente:2018xbv} which also address the $B$ meson decay anomalies, such as {\it e.g.}, Refs.~\cite{Sierra:2015fma, Belanger:2015nma, Altmannshofer:2016jzy, Celis:2016ayl, Cline:2017lvv, Ellis:2017nrp, Baek:2017sew, Fuyuto:2017sys, Cox:2017rgn, Falkowski:2018dsl, Darme:2018hqg, Singirala:2018mio, Baek:2018aru, Kamada:2018kmi, Guadagnoli:2020tlx} for $Z^\prime$ models, Refs.~\cite{Varzielas:2015sno, Cline:2017aed, Hati:2018fzc, Choi:2018stw, Datta:2019bzu} for leptoquark models, and Refs.~\cite{Bhattacharya:2015xha, Kawamura:2017ecz, Cline:2017qqu, Cerdeno:2019vpd, Barman:2018jhz, Darme:2020hpo} for models with one-loop solutions.

In the present paper, we propose to simultaneously solve all of the three above NP issues by constructing a class of models inspired by the model in Ref.~\cite{Cerdeno:2019vpd}, in which the DM was provided by a neutral $SU(2)_L$ singlet vectorlike fermion stabilized by a new $Z_2$ symmetry. By further introducing two extra scalar fields, one $SU(3)_c$ colored while the other colorless, the lepton universality violation observed in $B$ meson decays was solved by the NP one-loop contributions. We will extend this model by considering several simple variations of the $SU(2)_L \times U(1)_Y$ charge assignment of the newly introduced particles. Concretely, we will focus on models in which the $SU(2)_L$ representations of these particles are either singlet, doublet or triplet, and the vectorlike fermions have integer electric charge with values 0 or $\pm 1$. We will list all models satisfying these conditions and will identify the possible DM candidate in each model. After that, we will study in detail the DM and flavor phenomenology in one of the most promising models in this class. In our discussion, we will perform a scan in the parameter space of physical interest and identify regions which can solve the muon $g-2$ anomaly and the $B$ meson decay anomaly while providing a viable DM candidate.

The paper is organized as follows. In Sec.~\ref{ModelClass}, we extend the model in Ref.~\cite{Cerdeno:2019vpd} by listing all possible simple charge variations. Whenever possible we identify the DM candidate in each model. In the following sections, we discuss one specific model in this list. In Sec.~\ref{ModelDetail}, we write down the corresponding NP Lagrangian. The flavor observables are calculated analytically in Sec.~\ref{flavor}, including  the muon anomalous magnetic moment, $b\to s\mu^+\mu^-$, the mass difference in the $B_s$-$\bar{B}_s$ mixing, and $b\to s\gamma$. In Sec.~\ref{PhenoDM}, we address the DM phenomenology as predicted by this model. Specifically, we consider the constraints from DM relic density, DM direct detections and the invisible Higgs decay. In Sec.~\ref{ScanDM} we present the numerical results of our scan by taking all of the flavor and DM observables into account. Finally, conclusions and further discussions on collider signatures are given in Sec.~\ref{Conclusion}.

\section{A List of Possible Models}\label{ModelClass}
\begin{figure}[!ht]
\centering
\includegraphics[width = 0.5 \linewidth]{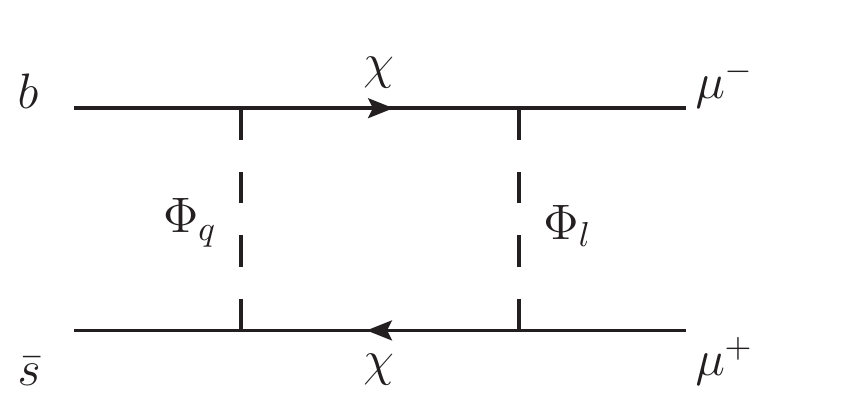}
\caption{One-loop Feynman diagram to solve the $R(K^{(*)})$ anomalies. }\label{Bloop}
\end{figure}

As discussed above we want to generalize the model proposed in Ref.~\cite{Cerdeno:2019vpd} by extending the dark sector particles to other $SU(2)_L \times U(1)_Y$ representations. 
%(\AM{What is written above gives the impression that $\chi$ can be electrically charged DM. Perhaps one should not call it dark fermion but instead \textit{new fermion} as I propose below.}) 
The classification is based on the representation of the new fermion $\chi$, belonging to the $Z_2$ odd sector, which either belongs to the singlet, doublet or triplet $SU(2)_L$ representation and has a $U(1)_Y$ hypercharge such that the electric charge is either 0, where it can become a DM candidate, or $\pm 1$. The charges of the remaining new fields can be determined from the existence of the following Yukawa interaction
\begin{eqnarray}\label{Yukawa}
{\cal L}^{\rm NP}_{\rm int} = y_{Q_i}\bar{Q}_{Li} \Phi_q \chi_R + y_{Li} \bar{L}_{Li} \Phi_l \chi_R +{\rm h.c.}\,,
\end{eqnarray}
where $\Phi_q$ and $\Phi_l$ are two spin zero fields, a triplet and a singlet of $SU(3)_c$ respectively;
 $y_{Q_i}$ and $y_{Li}$ are constants and  $Q_{Li}$ and $L_{Li}$ are the usual SM left-handed doublets 
 for quarks and leptons respectively.  
 This Lagrangian is required
 to provide the one-loop solution to the $B$ anomalies as shown in Fig.~\ref{Bloop}. 
%The chirality of $Q_{Li}$ and $L_{Li}$  have yielded by fitting the experimental $B$ meson decay data. 

We impose a new $Z_2$ symmetry under which the new particles $\Phi_q$, $\Phi_l$ and $\chi$ are all odd while all SM particles are even. With this charge assignment, the lightest color- and electromagnetically-neutral $Z_2$-odd particle can provide a DM candidate. Furthermore, we assume that $\chi$ is a vector-like fermion with its left- and right-handed components written as $\chi_L$ and $\chi_R$. However, note that only $\chi_R$ is involved in the interaction of Eq.~(\ref{Yukawa}). Note also that, if $\chi$ is in a self-conjugate representation of $SU(2)_L$ with zero $U(1)_Y$ charge, we only need to introduce the right-handed component $\chi_R$ since its Majorana mass term can arise in this case. This will be the case in two of the models presented. Otherwise $\chi_L$ has to be introduced.

In the remainder of this section, we list all possible NP models satisfying the restrictions above. Along with each model we identify the possible DM candidates. % rather than carefully studying its flavor and DM phenomenologies.
\begin{itemize}
\item Model 1:\\
The charge assignment for the $Z_2$-odd fields in this model is given in Table~\ref{Model1}, which is exactly the same as the one studied in Ref.~\cite{Cerdeno:2019vpd}. However, we would like to make some comments regarding this charge assignment. Since $\chi_R$ and the neutral component in $\Phi_l$ are electric- and color-neutral, either particle can be the DM candidate.
\begin{table}[!th]\caption{Charge Assignment for the $Z_2$-odd fields in Model 1}
{\begin{tabular}{c|ccc}
~ & $SU(3)_c$ & $SU(2)_L$ & $U(1)_Y$ \\
\hline
$\chi_R$ & {\bf 1} & {\bf 1} & 0\\
$\Phi_l$ & {\bf 1} & {\bf 2} & -1/2 \\
$\Phi_q$ & {\bf 3} & {\bf 2} & 1/6
\end{tabular}\label{Model1}}
\end{table}
Note that $\chi_R$ is self-conjugate, so we can introduce a Majorana mass term for it. Thus, if $\chi_R$ is the DM particle in this model, we do not need to introduce its left-handed partner $\chi_L$.  Furthermore, together with the following term
\begin{eqnarray}
(\Phi_l i\sigma_2 H)^2 + {\rm h.c.}\,,
\end{eqnarray}
where $H$ is the SM Higgs doublet, the Lagrangian breaks lepton number conservation by two units. Thus, the model can generate a non-zero Majorana neutrino mass at the one-loop level. In fact, this case is exactly the famous scotogenic model with the radiative neutrino mass generation proposed by Ernest Ma in Ref.~\cite{Ma:2006km}.

\item Model 2: \\
\begin{table}[!th]\caption{Charge Assignment for the $Z_2$-odd fields in Model 2}
{\begin{tabular}{c|ccc}
~ & $SU(3)_c$ & $SU(2)_L$ & $U(1)_Y$ \\
\hline
$\chi_R$ & {\bf 1} & {\bf 1} & 1\\
$\Phi_l$ & {\bf 1} & {\bf 2} & -3/2 \\
$\Phi_q$ & {\bf 3} & {\bf 2} & -5/6
\end{tabular}\label{Model2}}
\end{table}
Table~\ref{Model2} lists the quantum numbers for the fields in this model. Since there is no electrically neutral particle in the spectrum, 
there is no DM candidate, and we will no longer consider it.

\item Model 3:
\begin{table}[!th]\caption{Charge Assignment for the $Z_2$-odd fields in Model 3}
{\begin{tabular}{c|ccc}
~ & $SU(3)_c$ & $SU(2)_L$ & $U(1)_Y$ \\
\hline
$\chi_R$ & {\bf 1} & {\bf 1} & -1\\
$\Phi_l$ & {\bf 1} & {\bf 2} & 1/2 \\
$\Phi_q$ & {\bf 3} & {\bf 2} & 7/6
\end{tabular}\label{Model3}}
\end{table} \\
The charge assignment for the fields is given in Table~\ref{Model3}, where the DM candidate can only be the neutral component contained in the doublet scalar $\Phi_l$.

\item Model 4: \\
\begin{table}[!th]\caption{Charge Assignment for the $Z_2$-odd fields in Model 4}
{\begin{tabular}{c|ccc}
~ & $SU(3)_c$ & $SU(2)_L$ & $U(1)_Y$ \\
\hline
$\chi_R$ & {\bf 1} & {\bf 2} & 1/2\\
$\Phi_l$ & {\bf 1} & {\bf 1} & -1 \\
$\Phi_q$ & {\bf 3} & {\bf 1} & -1/3
\end{tabular}\label{Model4}}
\end{table}
Table~\ref{Model4} shows the charge assignment for this model, where the DM particle can only be the neutral component contained in the fermionic doublet $\chi$.

\item Model 5: \\
From the SM gauge group charges shown in Table~\ref{Model5}, we see that there are two DM candidates in this model: one is the singlet scalar $\Phi_l$ and the other is the neutral component 
of the vector-like fermion doublet $\chi$.
\begin{table}[!th]\caption{Charge Assignment for the $Z_2$-odd fields in Model 5}
{\begin{tabular}{c|ccc}
~ & $SU(3)_c$ & $SU(2)_L$ & $U(1)_Y$ \\
\hline
$\chi_R$ & {\bf 1} & {\bf 2} & -1/2\\
$\Phi_l$ & {\bf 1} & {\bf 1} & 0 \\
$\Phi_q$ & {\bf 3} & {\bf 1} & 2/3
\end{tabular}\label{Model5}}
\end{table}

\item Model 6:
\begin{table}[!th]\caption{Charge Assignment for the $Z_2$-odd fields in Model 6}
{\begin{tabular}{c|ccc}
~ & $SU(3)_c$ & $SU(2)_L$ & $U(1)_Y$ \\
\hline
$\chi_R$ & {\bf 1} & {\bf 3} & 0\\
$\Phi_l$ & {\bf 1} & {\bf 2} & -1/2 \\
$\Phi_q$ & {\bf 3} & {\bf 2} & 1/6
\end{tabular}\label{Model6}}
\end{table} \\
We show the SM gauge group charges of new $Z_2$-odd particles in Table~\ref{Model6}. Note that $\chi_R$ is self-conjugate, which guarantees the anomaly cancellation without the need of its left-handed component. Rather, we can introduce the Majorana mass term for $\chi_R$: $m_\chi \bar{\chi}^c_R \chi_R +{\rm h.c.}$. Thus, like in Model 1, small Majorana neutrino masses can be generated for active neutrinos~\cite{Ma:2008cu}.

\item Model 7:
\begin{table}[!th]\caption{Charge Assignment for the $Z_2$-odd fields in Model 7}
{\begin{tabular}{c|ccc}
~ & $SU(3)_c$ & $SU(2)_L$ & $U(1)_Y$ \\
\hline
$\chi_R$ & {\bf 1} & {\bf 3} & 1\\
$\Phi_l$ & {\bf 1} & {\bf 2} & -3/2 \\
$\Phi_q$ & {\bf 3} & {\bf 2} & -5/6
\end{tabular}\label{Model7}}
\end{table} \\
Table \ref{Model7} gives the charge assignment for this model, in which the only choice for DM is the neutral component in the triplet $\chi$.

\item Model 8:
\begin{table}[!th]\caption{Charge Assignment for the $Z_2$-odd fields in Model 8}
{\begin{tabular}{c|ccc}
~ & $SU(3)_c$ & $SU(2)_L$ & $U(1)_Y$ \\
\hline
$\chi_R$ & {\bf 1} & {\bf 3} & -1\\
$\Phi_l$ & {\bf 1} & {\bf 2} & 1/2 \\
$\Phi_q$ & {\bf 3} & {\bf 2} & 7/6
\end{tabular}\label{Model8}}
\end{table} \\
We show the color- and EW-charges of new particles in Table~\ref{Model8}. Here the DM can be the neutral component in either the scalar doublet $\Phi_l^0$ or the fermionic triplet $\chi^0$.
\end{itemize}

Finally, we would like to remark that if we swap the spins $0\leftrightarrow 1/2$ for new particles in the above models, we can generate other 8 models. For instance, if we exchange the spins of particles in Table~\ref{Model4}, the obtained model corresponds to the one studied in Ref.~\cite{Barman:2018jhz}.

\section{A Detailed Study of Model 5}\label{ModelDetail}
In order to study the connection between DM and flavor physics in the above class of models, we will now focus on model 5 in the following sections. As shown in Table~\ref{Model5}, there are three additional particles in this model: one fermion doublet $\chi_{L,R} = (\chi^0_{L,R}, \chi^-_{L,R})$ and two scalar $SU(2)_L$ singlets: $\Phi_l$ and $\Phi_q$, in which $\Phi_l$ is electrically neutral while $\Phi_q$ has colour and an electric charge of $2/3$. We also impose a $Z_2$ symmetry under which all new fields are odd while the SM fields are even under $Z_2$. As a result, we can introduce the following Dirac mass and Yukawa couplings to the fermion $\chi$:
\begin{eqnarray}\label{YukawaChi}
{\cal L} \supset m_\chi \bar{\chi}_L \chi_R + y_{Q_{i}} \bar{Q}_{L\, i} \Phi_q \chi_R + y_{L_{i}} \bar{L}_{L\,i} \Phi_l  \chi_R + {\rm h.c.}\,,
\end{eqnarray}
where $Q_{Li}$ and $L_{Li}$ denote the left-handed quark and lepton doublets in the SM. Note that both $\chi^0$ and $\chi^\pm$ share the same mass $m_\chi$ at tree level due to the fact that the Dirac mass term above is the sole source for both fermions.

We decompose the neutral complex scalar $\Phi_l$ into its real and imaginary components as $\Phi_l = (S+iA)/\sqrt{2}$. The scalar potential can be written as follows:
\begin{eqnarray}
V(H,\Phi_q,\Phi_l) &=& -\mu_H^2 |H|^2 +\mu^2_{\Phi_l} |\Phi_l|^2 +\mu_{\Phi_q}^2 |\Phi_q|^2 + \frac{\mu^{\prime\, 2}_{\Phi_l}}{2} (\Phi^2_l + \Phi^{*\, 2}_l) \nonumber\\
&& + \lambda_H |H|^4 + \lambda_{\Phi_q}|\Phi_q|^4 + \lambda_{\Phi_l} |\Phi_l|^ 4 \nonumber\\
&& + \lambda_{H\Phi_q} |H|^2 |\Phi_q|^2 + \lambda_{H\Phi_l} |H|^2 |\Phi_l|^2 + \lambda_{\Phi_q \Phi_l} |\Phi_q|^2 |\Phi_l|^2 \nonumber\\
&& + \frac{\lambda_{\Phi_l}^{\prime}}{4} (\Phi_l^2 + \Phi_l^{*\,2})^2 + \frac{\lambda^\prime_{\Phi_q \Phi_l}}{2} |\Phi_q|^2 (\Phi_l^2 + \Phi_l^{*\, 2} ) + \frac{\lambda^\prime_{H\Phi_l}}{2} |H|^2 (\Phi_l^2 + \Phi_l^{*\, 2} ) \,.
\end{eqnarray}
After electroweak symmetry breaking, the SM Higgs doublet $H$ acquires its vacuum expectation value $v_H = \sqrt{\mu_H^2/\lambda_H}$ and the scalar particles mass spectrum is given by
\begin{eqnarray}\label{ScalarMass}
m_h^2 = 2 \lambda_H v_H^2\,,&\quad& m_{\Phi_q}^2 = \mu_{\Phi_q}^2 + \frac{1}{2}\lambda_{H\Phi_q} v_H^2  \,,\nonumber\\
m_S^2 = \mu_{\Phi_l}^2 +\mu^{\prime \, 2}_{\Phi_l} + \frac{1}{2}(\lambda_{H\Phi_l}+\lambda^\prime_{H\Phi_l})v_H^2\,, &\quad& m_A^2 =  \mu_{\Phi_l}^2 -\mu^{\prime \, 2}_{\Phi_l} + \frac{1}{2}(\lambda_{H\Phi_l}-\lambda^\prime_{H\Phi_l})v_H^2\,,
\end{eqnarray}
where $h$ is the only component left in the SM Higgs doublet as $H = (0, (v_H+h)/\sqrt{2})^T$ in the unitary gauge. As argued in Sec.~\ref{ModelClass}, there are two potential DM candidates in this model: the neutral component $\chi^0$ in the fermionic doublet and the neutral scalar $S$ or the pseudoscalar $A$ in the scalar singlet $\Phi_l$. However, the fermionic candidate $\chi^0$ has a very large DM-nucleon scattering cross section due to the tree-level $Z$ mediation. In order to avoid the stringent experimental constraints from DM direct searches, such as XENON1T~\cite{XENON1t}, the fermionic DM mass should be pushed to be of ${\cal O}({\rm TeV})$. On the other hand, if all new particles are above 1 TeV,  the loop contributions to $b\to s\mu^+ \mu^-$ and $\Delta a_\mu$ are too small to solve the associated flavor anomalies, even if we tune the Yukawa couplings in Eq.~(\ref{YukawaChi}) to their perturbative limits $\sqrt{4\pi}$~\cite{Cerdeno:2019vpd}. Thus, we will not consider the fermionic DM  candidate $\chi^0$ and concentrate on the physics of the (pseudo)scalar DM $S(A)$. We should note that the DM and flavor phenomenology of this model is exactly the same for either $S$ or $A$. Hence, without loss of generality, we assume that $m_S < m_A$ so that $S$ comprises the whole DM density. According to Eq.~(\ref{ScalarMass}), it implies the following relation $\mu^{\prime \, 2}_{\Phi_l} + \frac12 \lambda^\prime_{H\Phi_l} v_H^2 <0$.

We can also rewrite the Yukawa interactions in Eq.~(\ref{YukawaChi}) relevant to solve the $B$ decay anomaly as follows:
\begin{eqnarray}
{\cal L} \supset y_{d_i} (  \bar{u}_{L\,j} V_{ji} \chi_R^0  + \bar{d}_{L\,i} \chi_R^- )\Phi_q + \frac{y_{e_i}}{\sqrt{2}}(S+iA) (\bar{\nu}_{L\,j} U_{ji} \chi_R^0 + \bar{e}_{L\,i} \chi_R^-) + {\rm h.c.}\,.
\end{eqnarray}
where we have defined the Yukawa couplings $y_{d_i (e_i)}$ which are obtained from $y_{Q_i (L_i)}$ by transforming the quarks and leptons into their mass eigenstates, and the matrix $V$ and $U$ are 
the Cabibbo-Kobayashi-Maskawa (CKM) and the Pontecorvo-Maki-Nakagawa-Sakata (PMNS) matrices, respectively. In order to suppress the strong flavor constraints on the first-generation quarks and leptons and to keep our discussion as simple as possible, we only allow these $Z_2$-odd particles to couple to quarks of the last two generations and the second-generation leptons. In other words, we only take $y_b$, $y_s$ and $y_\mu$ to be nonzero.

\section{Flavor Phenomenology}\label{flavor}
In this section we discuss the NP contributions to the various flavor observables in model 5. We will present the relevant analytic expressions 
used to perform the numerical scan presented in Sec.~\ref{ScanRes}.
\subsection{$(g-2)_\mu$}
The general amplitude of photon interactions with a charged particle can be written as
\begin{eqnarray}\label{defAMM}
\bar{u}(p^\prime) e\Gamma_\mu u(p) = \bar{u}(p^\prime) \Big[ e\gamma_\mu F_1(q^2) + \frac{ie\sigma_{\mu\nu}q^\nu}{2 m_f} F_2(q^2) + ... \Big]u(p)\,,
\end{eqnarray}
in which the photon momentum is defined as to flow into the vertex. The magnetic moment of muon is defined as $a_\mu = F_2(0)$. %Therefore, we should pick up the term with a single photon momentum $q$ in the amplitude. 
As mentioned in the Introduction, there is a long-standing discrepancy between the SM theoretical and experimental values of $a_\mu$ given in Eq.~(\ref{g2Val})~\cite{PDG,Gorringe:2015cma}. We hope to explain the anomalous magnetic moment of muon, i.e., $(g-2)_\mu$, within our model, where the leading-order contribution is provided by the one-loop diagrams enclosed by the negatively charged fermion $\chi_R$ and the neutral scalars $H$ or $A$. According to Ref.~\cite{Arnan:2016cpy}, the NP contribution is
\begin{eqnarray}\label{ExpG2}
\Delta a_\mu &=& \frac{m_\mu^2 |y_\mu|^2}{8\pi^2 m_\chi^2} \left(-\frac{1}{2} Q_\chi \right) \big[\tilde{F}_7(x_S) + \tilde{F}_7 (x_A)\big] \nonumber\\
&=& \frac{m_\mu^2 |y_\mu|^2}{16\pi^2 m_\chi^2} \big[\tilde{F}_7(x_S) + \tilde{F}_7 (x_A)\big]\,,
\end{eqnarray}
where
\begin{equation}\label{F7t}
\tilde{F}_7(x) = \frac{1-6x+3x^2+2x^3-6x^2 \ln x}{12(1-x)^4}\,,
\end{equation}
and $x_{S(A)} = m^2_{S(A)}/m_\chi^2$.

\subsection{$B\to K^{(*)}\mu^+\mu^-$}
It is easy to see that the anomalies in $B$ meson decays can be explained microscopically by the flavor-changing neutral current process $b\to s \mu^+ \mu^-$. In the present model, we can generate the following relevant effective Hamiltonian for $b \to s \mu^+ \mu^-$~\cite{Altmannshofer:2008dz, Becirevic:2012fy}:
\begin{eqnarray}
{\cal H}_{\rm eff} = -\frac{4 G_F}{\sqrt{2}}V_{tb}V_{ts}^* (C_9^{\rm NP} {\cal O}_9 +  C^{\rm NP}_{10} {\cal O}_{10})\,,
\end{eqnarray}
where
\begin{eqnarray}
{\cal O}_9 = \frac{\alpha}{4\pi}[\bar{s} \gamma^\nu P_L b][\bar{\mu}\gamma_\nu \mu]\,, \quad\quad {\cal O}_{10} = \frac{\alpha}{4\pi}[\bar{s} \gamma^\nu P_L b][\bar{\mu}\gamma_\nu \gamma^5 \mu]\,,
\end{eqnarray}
in which $\alpha$ is the fine structure constant of the electromagnetic interaction.

In our model, there are three kind of diagrams contributing to these two operators: box diagrams as well as $\gamma$- and $Z$-penguin diagrams. However, as shown in Ref.~\cite{Arnan:2016cpy}, the $Z$-penguin diagrams are suppressed by the factor $m_b^2/m_Z^2$ and can therefore be neglected. In what follows, we only consider the box and $\gamma$-penguin contributions.

The box diagrams in this model are shown in Fig.~\ref{Bloop} with the original complex scalar $\Phi_l$ replaced by its real and imaginary components, $S$ and $A$. They give new contributions to the Wilson coefficient $C^{\rm NP}_{9,10}$ as follows~\cite{Arnan:2016cpy}:
\begin{eqnarray}
C_9^{\rm box} = -C_{10}^{\rm box} = {\cal N} \frac{y_s y_b^* |y_\mu|^2}{64\pi \alpha m_\chi^2} [F(x_{\Phi_q}, x_S) + F(x_{\Phi_q}, x_A)]\,,
\end{eqnarray}
where $x_{\Phi_q, S, A} \equiv m_{\Phi_q, S, A}^2/m_\chi^2$ and ${\cal N}^{-1} = 4G_F V_{tb} V_{ts}^*/\sqrt{2}$. The function $F(x,y)$ is defined as
\begin{eqnarray}\label{FuncBox}
F(x,y) = \frac{1}{(1-x)(1-y)} + \frac{x^2 \ln x}{(1-x)^2(x-y)}+ \frac{y^2 \ln y}{(1-y)^2(y-x)}\,.
\end{eqnarray}

There are two $\gamma$-penguin diagrams differentiated by the internal lines from which the photon is emitted, since both loop particles, $\Phi_q$ and $\chi^-$, are electrically charged. Also, note that these two diagrams only generate the effective operator ${\cal O}_9$, with the corresponding Wilson coefficient given by\footnote{Compared with Eq.~(3.7) in Ref.~\cite{Arnan:2016cpy}, our result for $C_9^\gamma$ is larger by a factor of 2.}
\begin{eqnarray}
C_9^{\gamma} = {\cal N} \frac{y_s y_b^* }{m_\chi^2} [Q_{\Phi_q} F_9(x_{\Phi_q})-Q_\chi G_9(x_{\Phi_q})]\,,
\end{eqnarray}
where the functions $F_9(x)$ and $G_9(x)$ are defined by~\cite{Arnan:2016cpy}
\begin{eqnarray}
F_9(x) &=& \frac{-2x^3 +9x^2-18 x +11 + 6\ln x}{36(1-x)^4}\,, \nonumber\\
G_9(x) &=& \frac{7-36x+45x^2-16x^3+6(2x-3)x^2 \ln x}{36(1-x)^4}\,.
\end{eqnarray}

However, after numerical calculations of the box and $\gamma$-penguin diagrams, we find that the NP  amplitude of $b\to s\mu^+ \mu^-$ is always dominated by the box diagrams in our model, {\it i.e.}, $C_9^{\rm NP} = C_9^{\rm box}+ C_9^{\gamma} \simeq C_9^{\rm box} = -C_{10}^{\rm box} = -C^{\rm NP}_{10}$. This indicates that the relevant operator in our model reduces to a single left-handed one of the form $({\alpha}/{4\pi})[\bar{s} \gamma^\nu P_L b][\bar{\mu}\gamma_\nu (1-\gamma_5) \mu]$, which has been widely investigated in the literature~\cite{Capdevila:2017bsm,Descotes-Genon:2015uva, Hurth:2016fbr, Altmannshofer:2017yso, DAmico:2017mtc, Hiller:2017bzc, Geng:2017svp, Ciuchini:2017mik, Hurth:2017hxg, Alguero:2019ptt, Coy:2019rfr, Bhattacharya:2019dot, Vicente:2020usa, Biswas:2020uaq, Bhom:2020lmk, Alok:2017sui} because evidence for $R(K^{(*)})$ anomalies was observed in 2014. More recently, this scenario has been revisited in Ref.~\cite{Datta:2019zca} by fitting this single operator with the latest $b\to s\mu^+\mu^-$ and $R(K^{(*)})$ data measured by the LHCb and Belle Collaborations. The best fitted value of the Wilson coefficient is given by $C_9^{\rm NP} = -C^{\rm NP}_{10} = -0.53\pm 0,08$, with the improvement of the data fitting  by $5.8\sigma$ compared with the SM predictions. In our subsequent numerical scan of the parameter space, we only keep the models which can generate the Wilson coefficient $C_9^{\rm NP} = -C_{10}^{\rm NP}$ to be within the $2\sigma$ range around its central value. Note that recent works in Refs.~\cite{Datta:2019zca, Bhattacharya:2019dot} have shown that the single left-handed operator cannot provide a perfect fit to the whole set of $B$ meson decay data. In order to totally reduce the tension, one needs to consider extensions beyond this simple framework in the fits. However, we will not consider such complicated scenarios in the present work.

The rare decay process $B_s \to \mu^+ \mu^-$ may play a crucial role in constraining the present scenario with $C_9^{\rm NP} = -C_{10}^{\rm NP}$. In the SM, this decay channel is induced by the box and penguin diagrams. Due to helicity suppression of this process, only the operator ${\cal O}_{10}$ can contribute, with the SM expression of its branching fraction given by~\cite{Bobeth:2013uxa}:
\begin{eqnarray}\label{BsDecaySM}
{\cal B} (B_s \to \mu^+ \mu^-)^{\rm SM} = \tau_{B_s} f_{B_s}^2 m_{B_s}\frac{G_F^2 \alpha^2}{16\pi^3} |V_{tb} V_{ts}^*|  m_\mu^2  |C_{10}^{\rm SM}|^2 \sqrt{1-\frac{4 m^2_{\mu}}{m_{B_s}^2}} \,.
\end{eqnarray}
where $m_{B_s}$, $f_{B_s}$, and $\tau_{B_s}$ refer to the meson $B_s$'s mass, decay constant, and lifetime, respectively, and $C_{10}^{\rm SM}$ is the SM value to Wilson coefficient of the effective operator ${\cal O}_{10}$. On the other hand, our model can generate ${\cal O}_{10}$ via the NP box diagrams with its Wilson coefficient $C_{10}^{\rm NP}$. As a result, the NP contributions to this $B_s$ decay process is simply given by Eq.~({\ref{BsDecaySM}}) with the SM Wilson coefficient $C_{10}^{\rm SM}$ replaced by its NP one $C_{10}^{\rm NP}$~\cite{Barman:2018jhz}.

Numerically, the SM prediction of the branching ratio for $B_s \to \mu^+ \mu^-$ is given by~\cite{Bobeth:2013uxa}
\begin{eqnarray}
{\cal B} (B_s \to \mu^+ \mu^-)^{\rm SM} &=& (3.65 \pm 0.23) \times 10^{-9}\,,
\end{eqnarray}
while the measurement performed by the LHCb Collaboration has given~\cite{PDG, Aaij:2017vad}
\begin{eqnarray}
{\cal B}(B_s \to \mu^+ \mu^-)^{\rm Exp} &=& (2.7^{+0.6}_{-0.5}) \times 10^{-9}\,,
\end{eqnarray}
which shows that the measurement agrees with the SM value within $1\sigma$ confidence level (CL). In the following, we will constrain our model by requiring the NP contribution to this channel to be within the $2\sigma$ CL  experimentally allowed range.

\subsection{$B_s -\bar{B}_s$ Mixing}
A further important constraint on the parameter space related to the $b\to s$ transition is provided by the $B_s$-$\bar{B}_s$ mixing. Since the NP in our model only involves the left-handed SM fermions, the contribution to $B_s$-$\bar{B}_s$ mixing can only arise from the following single one effective operator
\begin{eqnarray}\label{BBHam}
{\cal H}^{B\bar{B}}_{\rm eff} = C_{B\bar{B}} Q_1 \equiv C_{B\bar{B}} (\bar{s}_\alpha \gamma^\mu P_L b_\alpha) (\bar{s}_\beta \gamma^\mu P_L b_\beta)\,,
\end{eqnarray}
where $\alpha$ and $\beta$ denote the color indices which are contracted in each pair. The NP contribution to the above Wilson coefficient in our model is given by~\cite{Arnan:2016cpy}
\begin{eqnarray}
C_{B\bar{B}}^{\rm NP} = \frac{(y_s y_b^*)^2}{128\pi^2 m_\chi^2} F(x_{\Phi_q}, x_{\Phi_q})\,,
\end{eqnarray}
where
\begin{eqnarray}
F(x,x) = \frac{1-x^2+2 x \ln x}{(1-x)^3}
\end{eqnarray}
is the function $F(x,y)$ defined in Eq.~(\ref{FuncBox}) in the limit of equal arguments.

The constraint is imposed on the mass difference $\Delta M_s$ between the two neutral meson states, $B_s$ and $\bar{B}_s$. According to Ref.~\cite{Arnan:2019uhr}, we can represent the constraint in terms of the ratio of the experimental value of the $B_s$ meson mass difference $\Delta M_s^{\rm exp}$ with its SM counterpart $\Delta M_s^{\rm SM}$ as follows~\cite{Arnan:2019uhr}:
\begin{equation}\label{RMsVal}
R_{\Delta M_s} = \frac{\Delta M_s^{\rm exp}}{\Delta M_s^{\rm SM}} - 1 = -0.09 \pm 0.08\,, \quad \mbox{ at $1\sigma$ C.L.} \,,
\end{equation}
where, in order to compute the SM result, we have used the value of the matrix element $\langle \bar{B}_s| Q_1 (\mu_b)|B_s\rangle$ obtained from a $N_f = 2+1$ lattice simulation in Ref.~\cite{Bazavov:2016nty}, which is consistent with the $N_f=2$ result in Ref.~\cite{Carrasco:2013zta}, the sum rules calculation in Ref.~\cite{King:2019lal}, and the most recent FLAG-2019 lattice average value in Ref.~\cite{Aoki:2019cca}. Here $Q_1(\mu_b)$ is the effective operator defined in Eq.~(\ref{BBHam}) at the scale $\mu_b$. If we further identify $\Delta M_s^{\rm exp}$ as the total contribution to the $B_s$-$\bar{B}_s$ mixing mass difference, we can write the quantity $R_{\Delta M_s}$ in terms of the NP and SM Wilson coefficients as follows~\cite{Arnan:2019uhr, Gabbiani:1996hi}:
\begin{eqnarray}
R_{\Delta M_s} = \left|1+ \frac{0.8 C_{B\bar{B}}^{\rm NP}(\mu_H)}{C_{B\bar{B}}^{\rm SM}(\mu_b)} \right|-1\,,
\end{eqnarray}
where $C_{B\bar{B}}^{\rm NP} (\mu_H)$ is the NP Wilson coefficient defined at a high-energy scale of $\mu_H = 1$~TeV, and $C^{\rm SM}_{B\bar{B}} \simeq 7.2 \times 10^{-11} \, {\rm GeV}^{-2}$ is the corresponding SM value defined at the scale $\mu_b$ computed by employing the results in Ref.~\cite{Bazavov:2016nty}. Also, the factor 0.8 is caused by the renormalization group running and the operator mixing as the scale decreases from $\mu_H$ to $\mu_b$.

Note that it is easily seen from Eq.~(\ref{RMsVal}) that there is a little tension between experimental measurements and the SM prediction as pointed out in Refs.~\cite{DiLuzio:2018wch,DiLuzio:2017fdq}. However, we did not try to solve this tension in the present paper. Rather, we will constrain $C_{B\bar{B}}^{\rm NP}$ by requiring the $R_{\Delta M_s}$ to lie in its $2\sigma$ confidence interval.

\subsection{$b \to s\gamma $}
Another strong constraints on our model arises from the $b\to s\gamma$ processes. The relevant effective Hamiltonian is given by~\cite{Arnan:2016cpy}
\begin{eqnarray}\label{bsgHam}
{\cal H}^{\gamma}_{\rm eff} = -\frac{4G_F}{\sqrt{2}} V_{tb} V_{ts}^* (C_7 {\cal O}_7 + C_8 {\cal O}_8)\,,
\end{eqnarray}
with
\begin{eqnarray}\label{bsgO}
{\cal O}_7 =\frac{e}{16\pi^2} m_b \bar{s}\sigma^{\mu\nu}P_R b F_{\mu\nu}\,, \quad {\cal O}_8 = \frac{g_s}{16\pi^2} m_b \bar{s}_\alpha \sigma^{\mu\nu} P_R T^a_{\alpha \beta} b_\beta G_{\mu\nu}^a\,,
\end{eqnarray}
where $F_{\mu\nu}$ and $G_{\mu\nu}^a$ stand for the field strength tensors for photons and gluons, respectively. Note that even though ${\cal O}_8$ cannot give direct contributions to $b\to s\gamma$, it would affect the final result via its mixing with ${\cal O}_7$ as the renormalization scale decreases.

In our model, the leading-order contribution to $b\to s\gamma$ is given at one-loop order, leading to the following Wilson coefficients for ${\cal O}_7$ and ${\cal O}_8$~\cite{Arnan:2016cpy}
\begin{eqnarray}
C_7 &=& {\cal N} \frac{y_s y_b^*}{2m_\chi^2} \left[Q_{\Phi_q} F_7(x_{\Phi_q})-Q_\chi \tilde{F}_7(x_{\Phi_q})\right]\,,\nonumber\\
C_8 &=& {\cal N} \frac{y_s y_b^*}{2m_\chi^2} F_7(x_{\Phi_q})\,,
\end{eqnarray}
where
\begin{eqnarray}
F_7 (x) = \frac{2+3x-6x^2+x^3+6x\ln x}{12(1-x)^4}\,,
\end{eqnarray}
while $\tilde{F}_7(x)$ has been shown in Eq.~(\ref{F7t}).

Currently, the most precise experimental measurement on the branching ratio of $b\to s\gamma$ is given by the HFAG Collaboration~\cite{Amhis:2016xyh}:
\begin{eqnarray}\label{bsgEx}
{\cal B}^{\rm exp}(b\to s\gamma) = (3.32 \pm 0.15)\times 10^{-4}\,,
\end{eqnarray}
while the SM prediction of the branching ratio for this process is~\cite{Misiak:2015xwa,Misiak:2017woa}
\begin{eqnarray}\label{bsgSM}
{\cal B}^{\rm SM}(b\to s\gamma) = (3.36 \pm 0.23) \times 10^{-4}\,,
\end{eqnarray}
which shows a good agreement between the experiments and theoretical calculations. In order to impose the $b \to s\gamma$ constraint on our model, we follow Ref.~\cite{Arnan:2016cpy, Arnan:2019uhr} to define
\begin{eqnarray}
R_{s\to \gamma} = \frac{{\cal B}^{\rm tot}(b\to s\gamma)}{{\cal B}^{\rm SM}(b\to s\gamma)} - 1 = -2.87(C_7 + 0.19 C_8)\,,
\end{eqnarray}
where ${\cal B}^{\rm tot}(b \to s\gamma)$ refers to the total branching ratio of $b\to s \gamma$ in our model including the NP contribution. Here the combination $C_7 + 0.19 C_8$ accounts for the mixing effect between effective operators ${\cal O}_7$ and ${\cal O}_8$ due to the renormalization group running from QCD calculations~\cite{Misiak:2015xwa,Misiak:2017woa}. On the other hand, by appropriately combining the experimental and theoretical errors in Eqs.~(\ref{bsgEx}) and (\ref{bsgSM}), it can be shown $R_{b\to s\gamma} = (-0.7 \pm 8.2)\times 10^{-2}$ at the $2\sigma$ confidence level~\cite{Arnan:2019uhr}, which can be transformed into
\begin{eqnarray}
| C_7 + 0.19 C_8 | \lesssim 0.06 \quad \mbox{at $2\sigma$ C.L.}\,.
\end{eqnarray}

\section{Dark Matter Phenomenology}\label{PhenoDM}
As discussed, the neutral scalar component $S$ contained in the singlet $\Phi_l$ is the lightest $Z_2$-odd particle. It is therefore stable and can play the role of DM candidate. In what follows, we will discuss the DM phenomenology, by exploring the DM relic density and constraints from DM searches.

\subsection{Dark Matter Relic Density}
Since $S$ is the only DM candidate it should reproduce the observed DM relic abundance. Currently, the most accurate measurement of this important quantity is provided by the Planck Collaboration with $\Omega_{\rm DM} h^2 = 0.1199\pm 0.0022$~\cite{Ade:2015xua}. Here we assume that the DM relic density is generated by the ordinary freeze-out mechanism, so that the relic abundance of $S$ can be determined by solving the following Boltzmann equation:
\begin{eqnarray}\label{DMBE}
\frac{d n_S}{dt} +3 H n_S = -\langle  \sigma v \rangle (n_S^2 -n_S^{\rm eq\,2})\,,
\end{eqnarray}
where $n_S$ denotes the number density of $S$ with $n^{\rm eq}$ as its corresponding equilibrium value, $H$ is the Hubble parameter and $\langle \sigma v \rangle$ refers to the thermal average of the DM annihilation cross section times the relative velocity' $v$. 

\begin{figure}[!ht]
\centering
\includegraphics[width = 0.8 \linewidth]{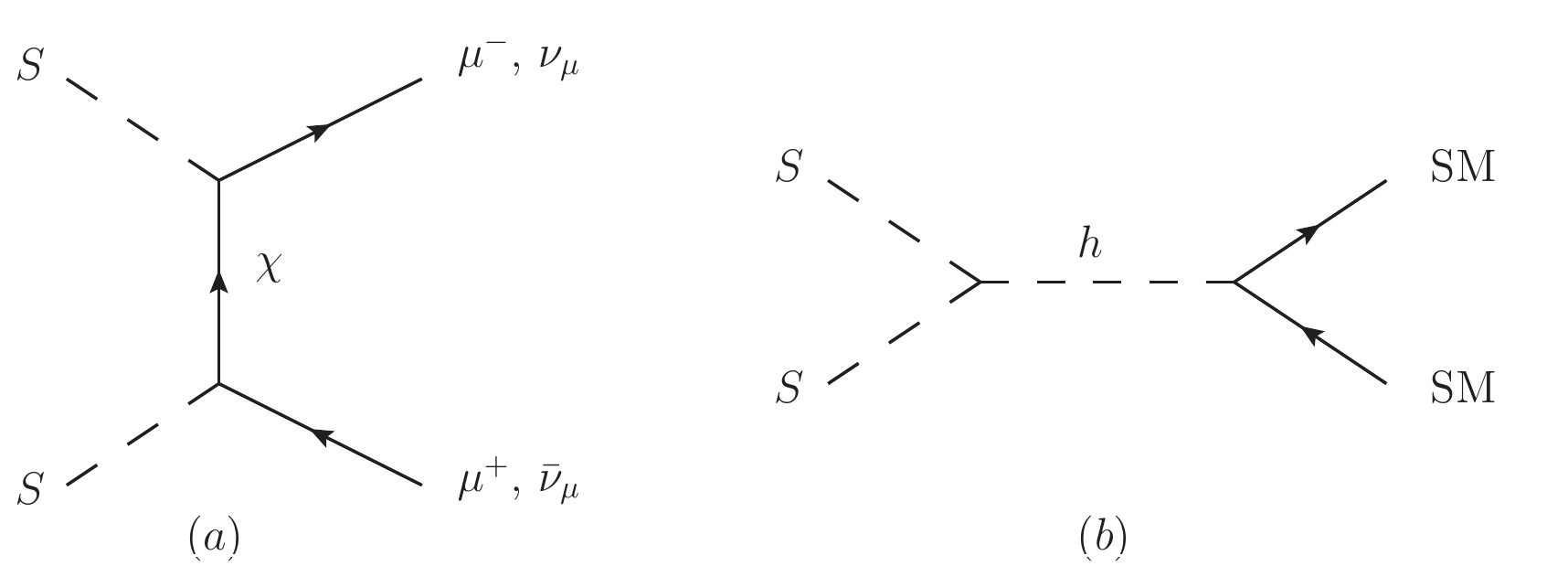}
\caption{Feynman diagrams for DM $S$ annihilations: (a) $SS \to \mu^+ \mu^-$ via the $\chi$ mediation; (b) $SS \to {\rm SM\, SM}$ via the SM-like Higgs mediation in which ``SM'' represents all massive SM particles.}\label{Sann}
\end{figure}

The two main classes of DM annihilation processes crucial to determine the DM relic abundance are presented in Fig.~\ref{Sann}. On the left we show $S$ pair annihilation into a $\mu^+ \mu^-$ ($\nu_\mu \bar{\nu}_\mu$) pair via the $t$- and $u$-channel $\chi^-$ ($\chi^0$) mediation. On the right, the $s$-channel annihilation mode mediated by the SM Higgs $h$ is shown.  
The process $SS \to \mu^+ \mu^-$ is dominated by the $d$-wave contribution in the zero muon mass limit, with its cross section given by
\begin{eqnarray}\label{SAnn1}
\langle \sigma v \rangle_{SS\to \mu^+\mu^-} = \frac{|y_\mu|^4}{240\pi} \frac{m_S^6}{(m_\chi^2 + m_S^2)^4} \langle v^4 \rangle = \frac{|y_\mu|^4}{128\pi} \frac{m_S^6}{(m_\chi^2 + m_S^2)^4}  \frac{1}{x^2}\,,
\end{eqnarray}
where the angle bracket refers to taking the thermal average of the corresponding quantity, and we have used the formula $\langle v^4 \rangle = 15/(8 x^2)$ for the non-relativistic Boltzmann distribution~\cite{Gondolo:1990dk} in which $x^{-1}\equiv T/m_S \approx 1/25$ with $T$ being the plasma temperature at the DM freeze-out time in the Universe. The cross section
for the process $SS\to \nu_\mu \bar{\nu}_\mu$ is also given by Eq.~(\ref{SAnn1}). For the DM annihilation processes with the $h$ mediation, the cross sections for different final states are all proportional to the Higgs portal coupling $(\lambda_{H\Phi_l}+\lambda_{H\Phi_l}^\prime)$, which are strongly constrained by DM direct detection results. 
\begin{figure}[!ht]
\centering
\includegraphics[width = 0.8 \linewidth]{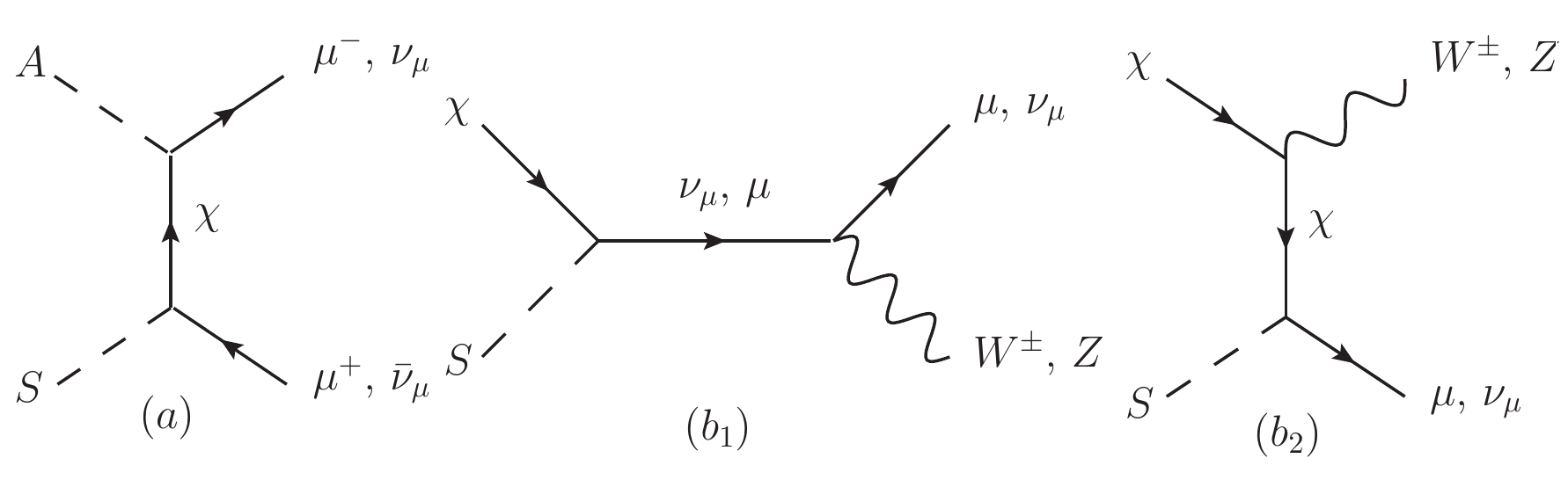}
\caption{Illustration of Feynman diagrams for (a) $SA$ and (b) $S\chi$ co-annihilation processes relevant to the determination of the DM relic abundance. }\label{Sco}
\end{figure}
Using several benchmark sets of parameters we found that when the mass difference between $S$ and $A$($\chi$) is comparable to or smaller than the temperature of the Universe, the number density of $A$ ($\chi$) is abundant at the DM freeze-out, and the co-annihilation $SA$($S\chi$) channels illustrated in Fig.~\ref{Sco} are still active in determining the model prediction of the DM relic density.

In our work, we numerically solve the Boltzmann equation in Eq.~(\ref{DMBE}) by taking advantage of the modified {\tt MicrOMEGAs v4.3.5} code~\cite{Belanger:2006is, Belanger:2014vza} which takes all possible co-annihilation channels into account.
\begin{figure}[!ht]
\centering
\includegraphics[width = 0.5 \linewidth]{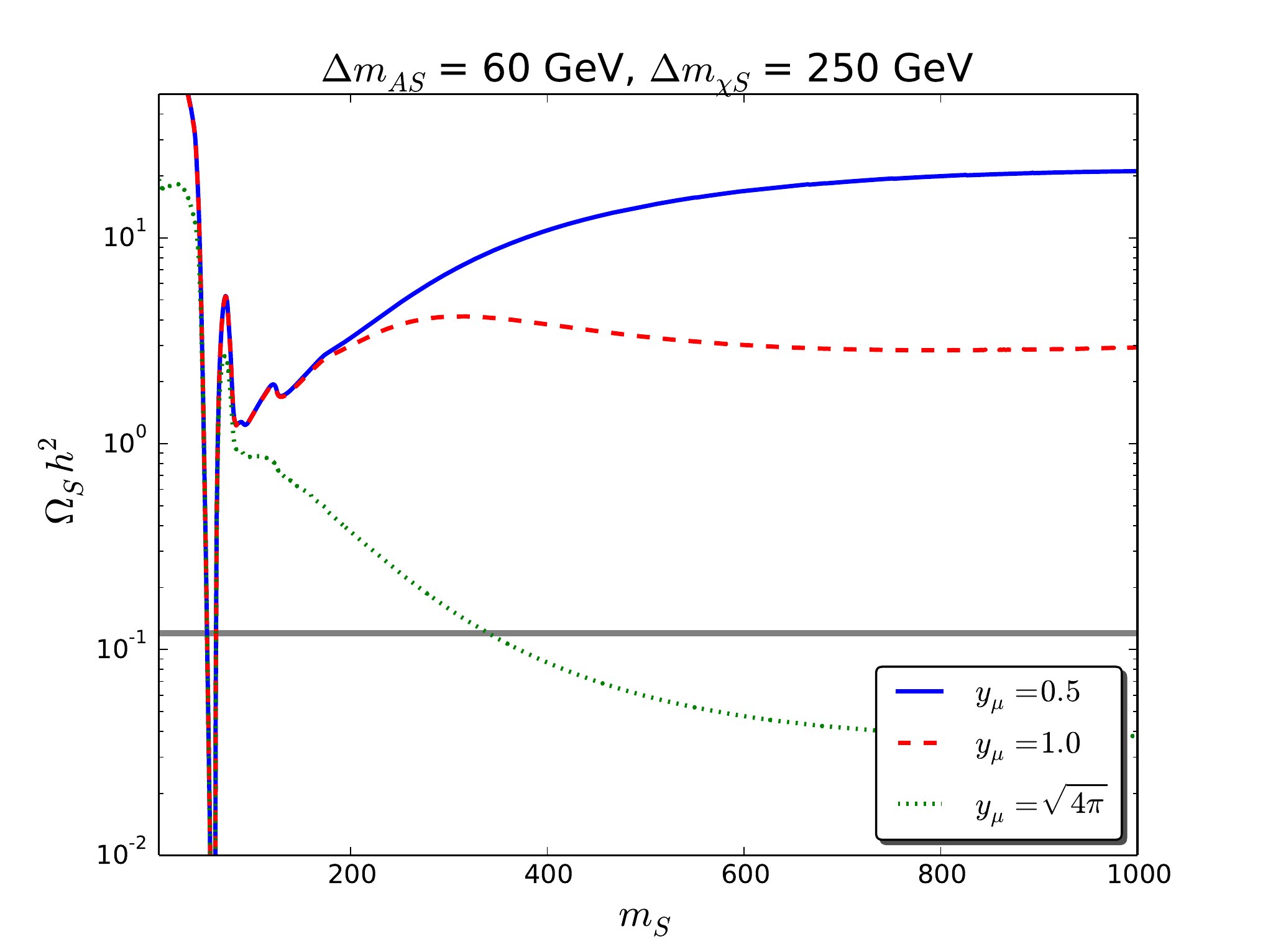}
\caption{The DM $S$ relic density as a function of the DM mass $m_S$. Here we fix the difference between the particle $A$ ($\chi$) mass and the DM mass to be $\Delta m_{AS} \equiv m_A-m_S = 60$~GeV ($\Delta m_{\chi S} \equiv m_\chi-m_S = 250$~GeV). Other relevant parameters are fixed to be $m_{\Phi_q} = 2000$~GeV, $\lambda_{H\Phi_l} = \lambda_{H\Phi_l}^\prime = 5\times 10^{-3}$, and $y_s = -y_b = 0.1$. }\label{mS}
\end{figure}
As an example of our numerical calculation, we show in Fig.~\ref{mS} the variation of the DM relic abundance as a function of the DM mass $m_S$ for different leptonic Yukawa couplings $y_\mu = 0.5$, 1.0, and $\sqrt{4\pi}$ with the last value corresponding to the perturbative limit. We have fixed the relevant parameters to be $m_{\Phi_q} = 2$~TeV, $\lambda_{H\Phi_l} = \lambda_{H\Phi_l}^\prime = 5\times 10^{-3}$ and $y_s = -y_b = 0.1$, as well as the mass differences to be $\Delta m_{AS} \equiv m_A- m_S = 60$~GeV and $\Delta m_{AS} \equiv m_A- m_S = 250$~GeV. From Fig.~\ref{mS}, it is seen that when $(\lambda_{H\Phi_l} + \lambda_{H\Phi_l}^\prime )$ is $10^{-2}$ and $y_\mu$ is small, {\it i.e.}, $y_\mu \lesssim 0.5$, the DM relic abundance is only satisfied in the region near the SM Higgs resonance $m_S \simeq m_h/2$, and the dominant DM annihilation proceeds via the $s$-channel SM Higgs mediation. When $y_\mu$ is increased to be around 1, the $d$-wave suppressed process $SS \to \mu^+ \mu^-$, $\nu_\mu \bar{\nu}_\mu$ induced by the new Yukawa interactions begins to be comparable to and even dominant over the SM Higgs mediated diagrams in the high DM mass region. However, the total DM annihilation cross section is still insufficient to lower the DM relic abundance to its experimentally allowed values. Finally, when $y_\mu$ becomes even larger, up to the perturbative limit $\sqrt{4\pi}$, a second allowed DM region appears in the high DM mass region where the $t$- and $u$-channel $S$ annihilations into leptons dominate the DM freeze-out, over the whole range of DM masses except for the Higgs resonance region.

\subsection{Constraints from Dark Matter Direct Detection and Higgs Invisible Decays}
In this subsection, we will focus on the experimental constraints from various DM searches. Let us begin by discussing DM direct detection, which may place severe constraints on the spin-independent DM-nucleon scattering. In the present model, the dominant DM direct detection signal arises through the tree-level diagram with $t$-channel SM-like Higgs mediation, leading to the following DM-nucleon scattering cross section
\begin{eqnarray}\label{DMDD}
\sigma(S N \to SN) = \frac{(\lambda_{H\Phi_l}+\lambda_{H\Phi_l}^\prime)^2}{4\pi} \frac{f_N^2 m_N^2 \mu_{SN}^2}{m_S^2 m_h^4}\,,
\end{eqnarray}
where $f_N \simeq 0.3$ denotes the effective Higgs-nucleon coupling~\cite{Cline:2013gha, Alarcon:2011zs, Ling:2017jyz}, $m_N$ is the nucleon mass, and   $\mu_{SN} \equiv m_S m_N/(m_S+m_N)$ is the reduced mass of the DM-nucleon system. At present, the best experimental upper bound on the DM direct detection cross section for a mass above $6$~GeV is provided by the XENON1T experiment~\cite{XENON1t}, which will be taken into account in our scan.

Collider searches impose further restrictions on dark matter. These are particularly relevant when $m_S < m_h/2$ because the DM particle $S$ is subject to the constraint from the SM-like Higgs boson invisible decay into an $S$ pair. The invisible decay width in our model is given by
\begin{eqnarray}\label{DMInv}
\Gamma(h\to SS) = \frac{(\lambda_{H\Phi_l}+\lambda_{H\Phi_l}^\prime)^2 v_H^2}{32\pi m_h} \sqrt{1-\frac{4m_S^2}{m_h^2}}\,.
\end{eqnarray}
Currently, the upper bound on this process is provided by LHC with ${\cal B}(h\to SS) \leq 0.24$~\cite{PDG}. Note that both the DM direct detection signal in Eq.~(\ref{DMDD}) and the SM-like Higgs invisible decay in Eq.~(\ref{DMInv}) only depend on two parameters: $m_S$ and $(\lambda_{H\Phi_l} + \lambda_{H\Phi_l}^\prime)$. As a result, the constraint from the Higgs invisible decay is always weaker than that of DM direct detection in the parameter space of interest.

%%% Paragraph for DM indirect detection %%%%%%%%%%%%%%%

\section{Results}\label{ScanRes}
In this section we discuss the results obtained by analysing the flavor and DM physics constraints in our model. We perform a multi-parameter scan to find out the common parameter regions that can satisfy all relevant flavor constraints: $R(K^{(*)})$, ${\cal B}(B_s \to \mu^+ \mu^-)$, $B_s$-$\bar{B}_s$ mixings, $b\to s\gamma$, the muon anomalous magnetic moment $\Delta a_\mu$, and the DM constraints, where the latter means both the correct DM relic abundance and the bounds on the direct detection searches. In order to simplify our analysis, we make the following restrictions of our parameter space. As shown in the formulae related to the $B_s$ meson decays and the $B_s$-$\bar{B}_s$ mixing, only the combination $y_s y_b^*$ appears. Also, in order to solve the deficit observed in the  measurements of $R(K^{(*)})$, this combination should be negative. Therefore, in our numerical scan we take $y_s$ and $y_b$ to be real with $y_s = -y_b/4$. Regarding DM, since the Higgs portal coupling is always of the form $(\lambda_{H\Phi_l}+\lambda_{H\Phi}^\prime)$, we take $\lambda_{H\Phi} =\lambda_{H\Phi}^\prime$. 
Moreover, the singlet $\Phi_q$ can be pair produced via gluon/quark fusion at the LHC, and one of its main decay channels is an up-type quark plus a $\chi^0$ which in turn decays into the DM particle $S$ and a neutrino, leading to a dijet plus missing transverse energy final state, {\it i.e.} $ jj +\slashed{E}_T$. According to a similar study in Ref.~\cite{Cerdeno:2019vpd}\footnote{Although the colored scalar $\Phi_q$ in Ref.~\cite{Cerdeno:2019vpd} has a different electroweak quantum number from the one in the present paper, the main production channel is still through the QCD processes. Thus, the constraint on $\Phi_q$ can be directly applied in our case.}, the lower limit on $\Phi_q$ is around $1$~TeV. Therefore, in order to avoid such a strong constraint, we fix the mass of the colored scalar $\Phi_q$ to be $m_{\Phi_q} = 1.5$~TeV. Furthermore, for $S$ to be the DM candidate, we require all other particles in the dark sector, including $A$ and $\chi$, to be heavier than $S$ by at least $10$~GeV, but all these particles should be lighter than 1~TeV. A further constraint coming from LEP searches for unstable heavy vector-like charged leptons~\cite{Achard:2001qw}, which sets a lower limit on the mass of the charged fermion $\chi^\pm$ of 101.2~GeV. We also impose this limit in our scan. For dimensionless couplings, we allow $(\lambda_{H\Phi}+\lambda_{H\Phi}^\prime)\leqslant 1$, $|y_b|\leqslant 1$, and $0 \leqslant y_\mu \leqslant \sqrt{4\pi}$.

\begin{figure}[h!]
	\centering
	\hspace{-5mm}
	\includegraphics[scale=0.45]{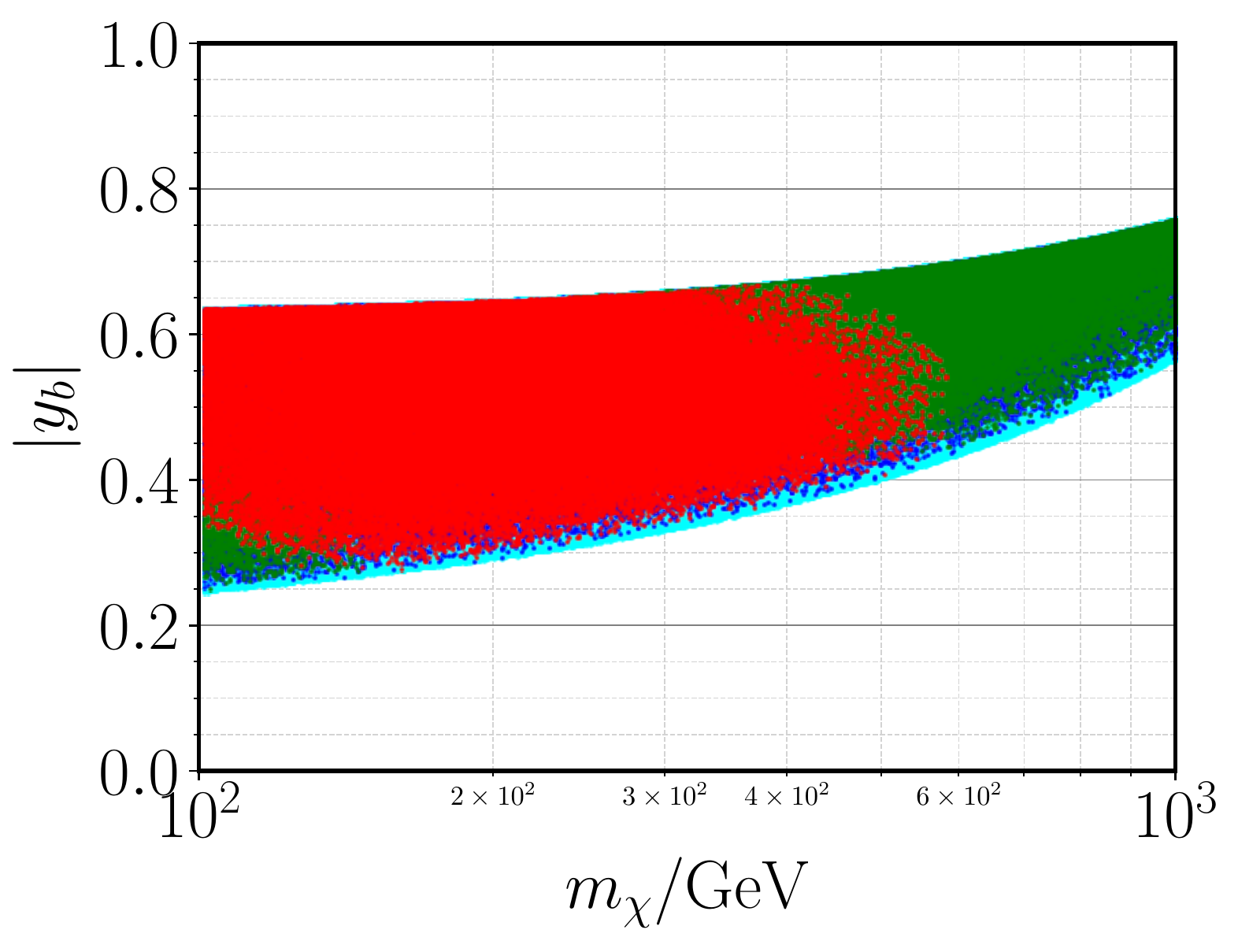}
	\includegraphics[scale=0.45]{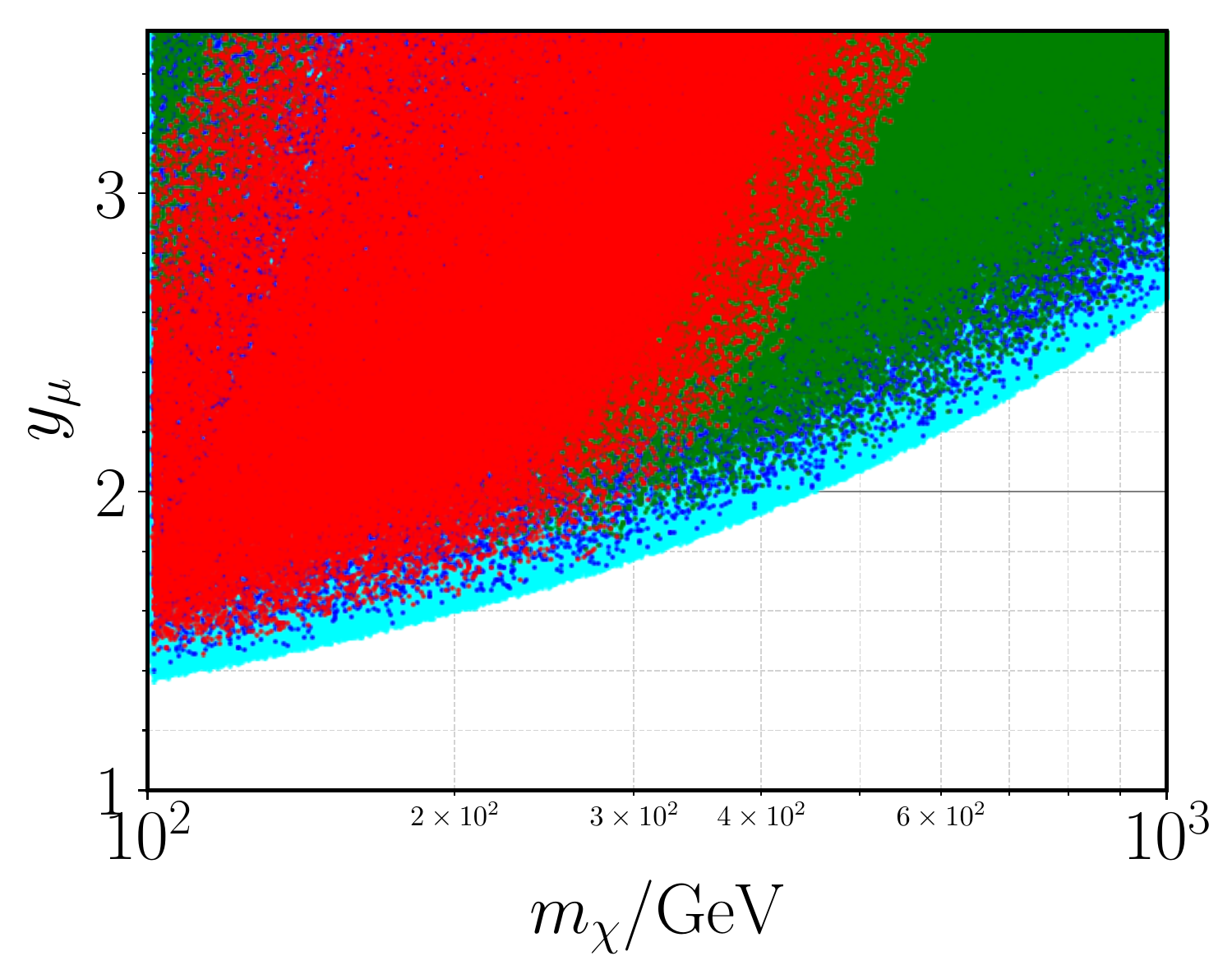}
	\includegraphics[scale=0.45]{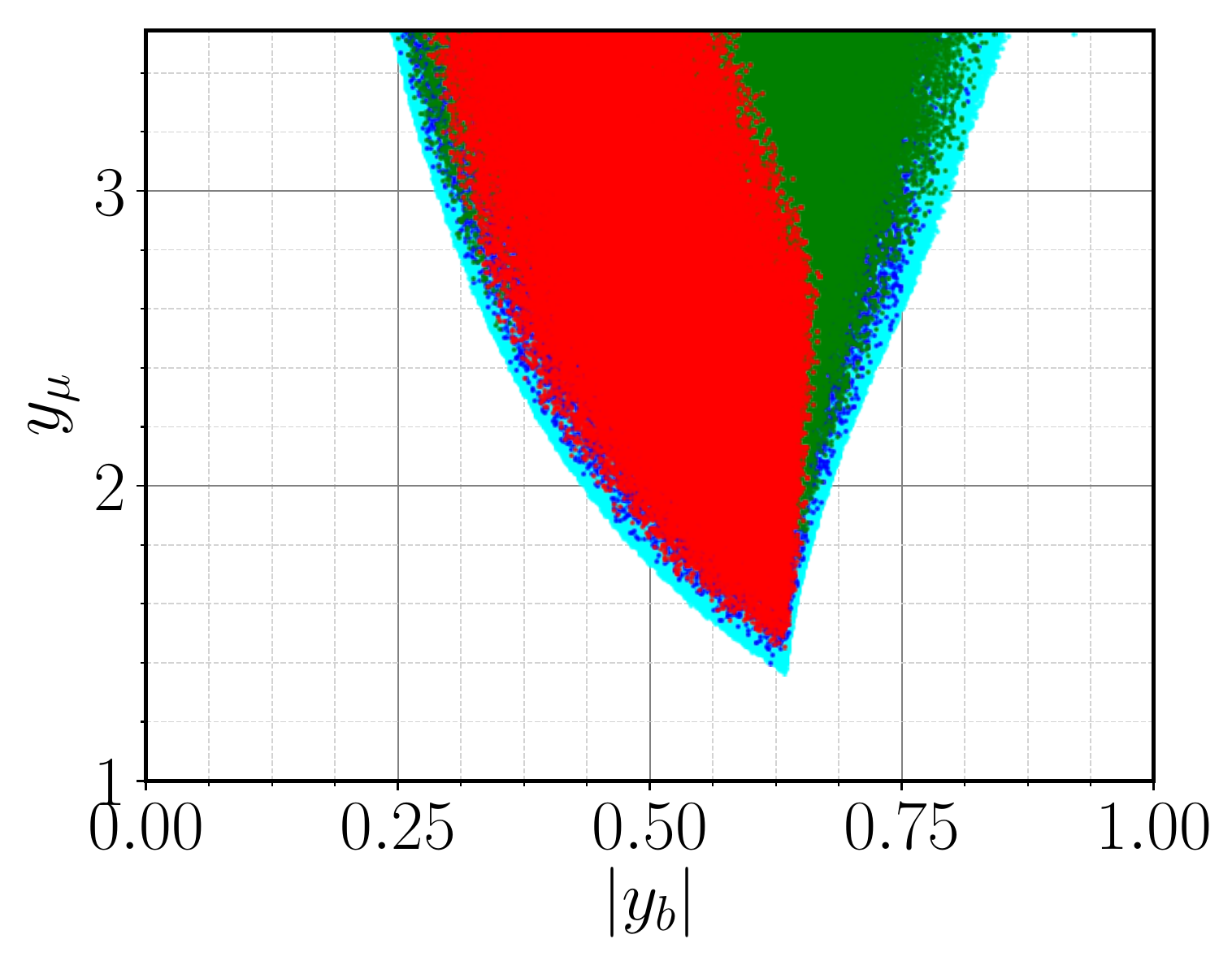}	
	\includegraphics[scale=0.45]{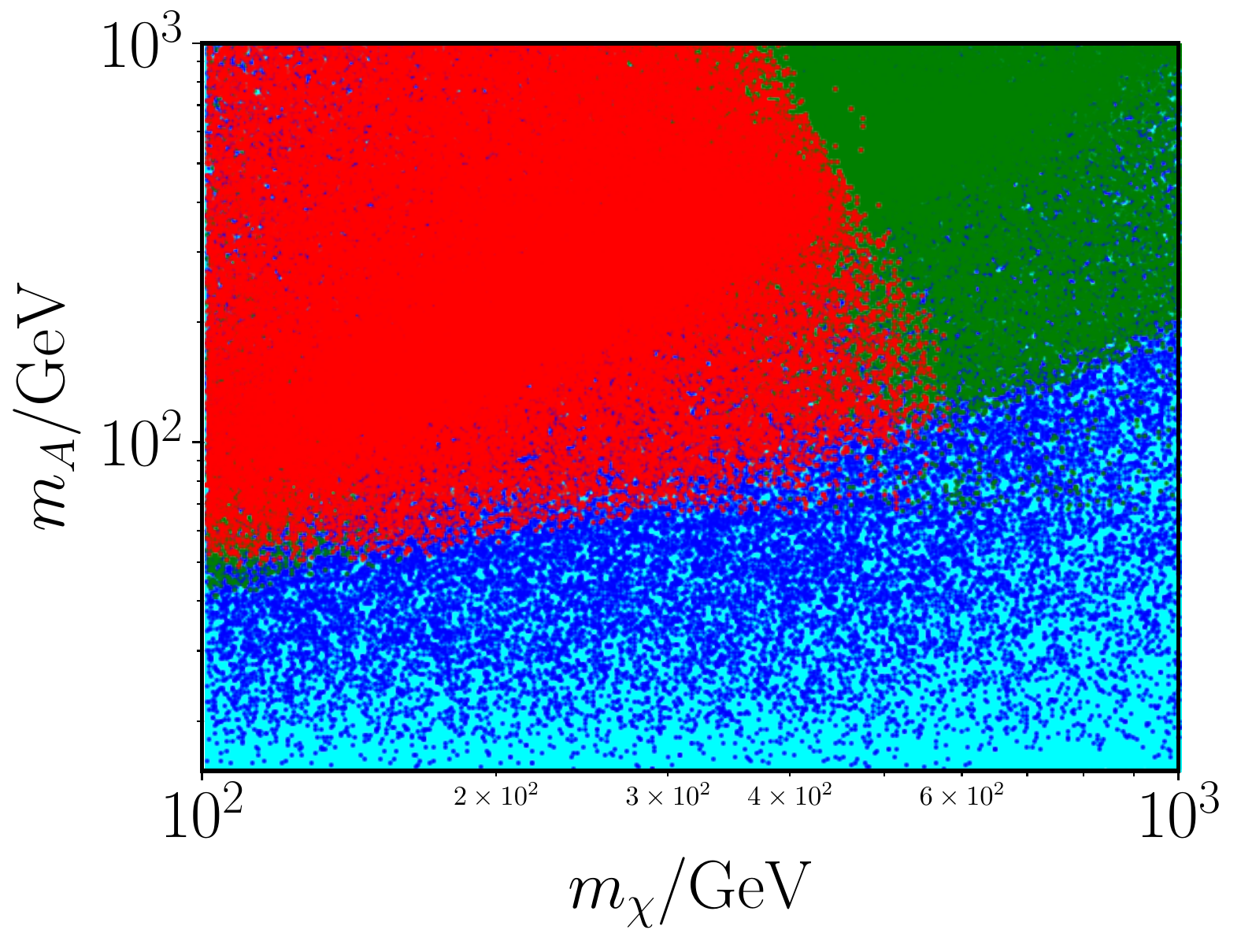}		
	\caption{Allowed parameter space projected in the planes $m_\chi$-$|y_b|$ (Top Left), $m_\chi$-$y_\mu$ (Top Right), $|y_b|$-$y_\mu$ (Bottom Left) and $m_\chi$-$m_A$ (Bottom Right). In the scan, we have fixed $m_{\Phi_q} = 1.5$~TeV and chosen $y_s = -y_b/4$ and $\lambda_{H\Phi_l} = \lambda_{H\Phi_l}^\prime$. All points satisfy the constraints from the non-anomalous B-physics results.
We shown in cyan the points that explain $R(K^{(*)})$; the blue points explain  $R(K^{(*)})$ and the DM relic density; the green points explain the B anomalies and DM relic density while satisfying all
constraints except the muon $(g-2)$; red points satisfy all constraints.	
 } \label{ScanFlavor}
\end{figure}

\begin{figure}[h!]
	\centering
	\hspace{-4mm}
	\includegraphics[scale=0.45]{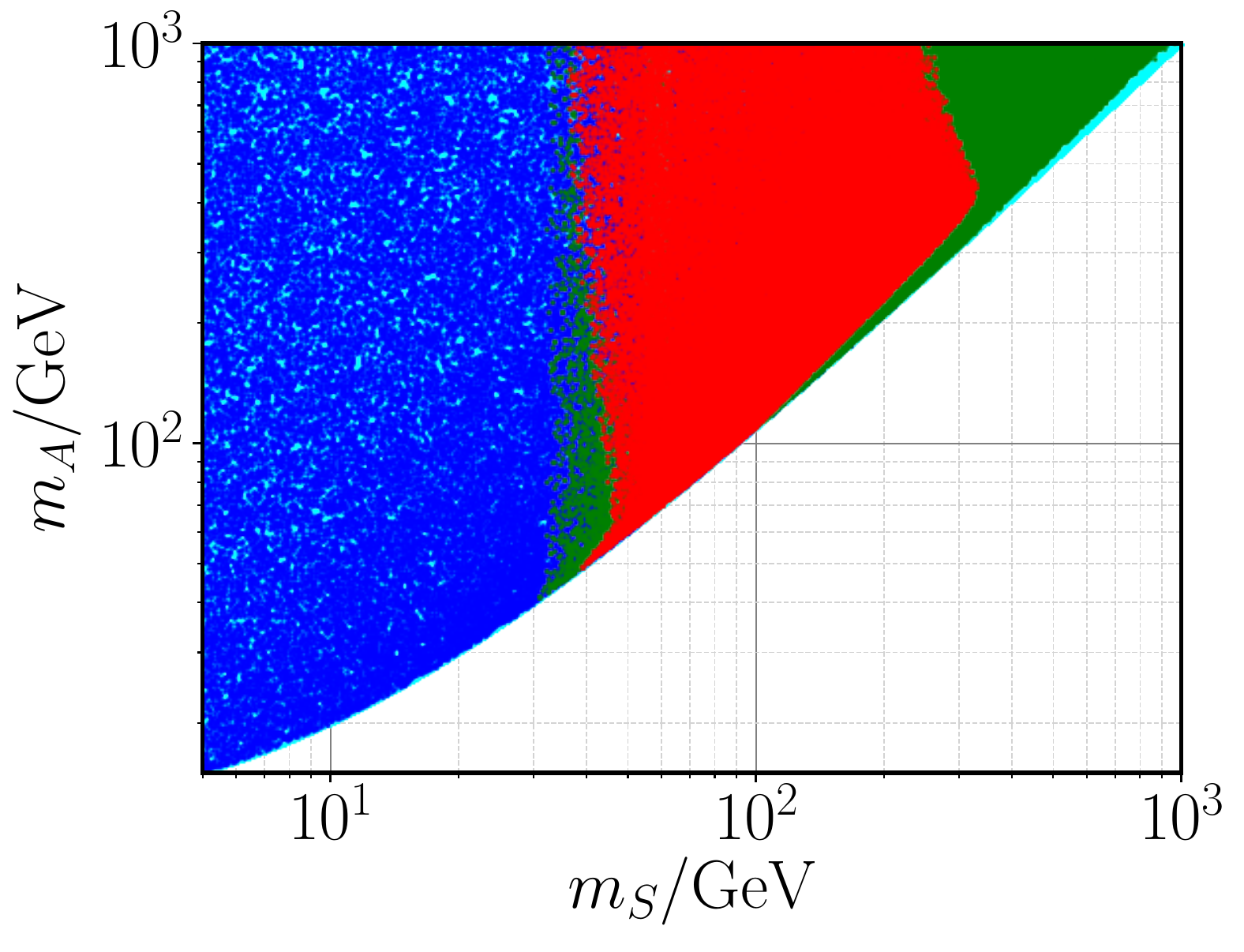}
	\includegraphics[scale=0.45]{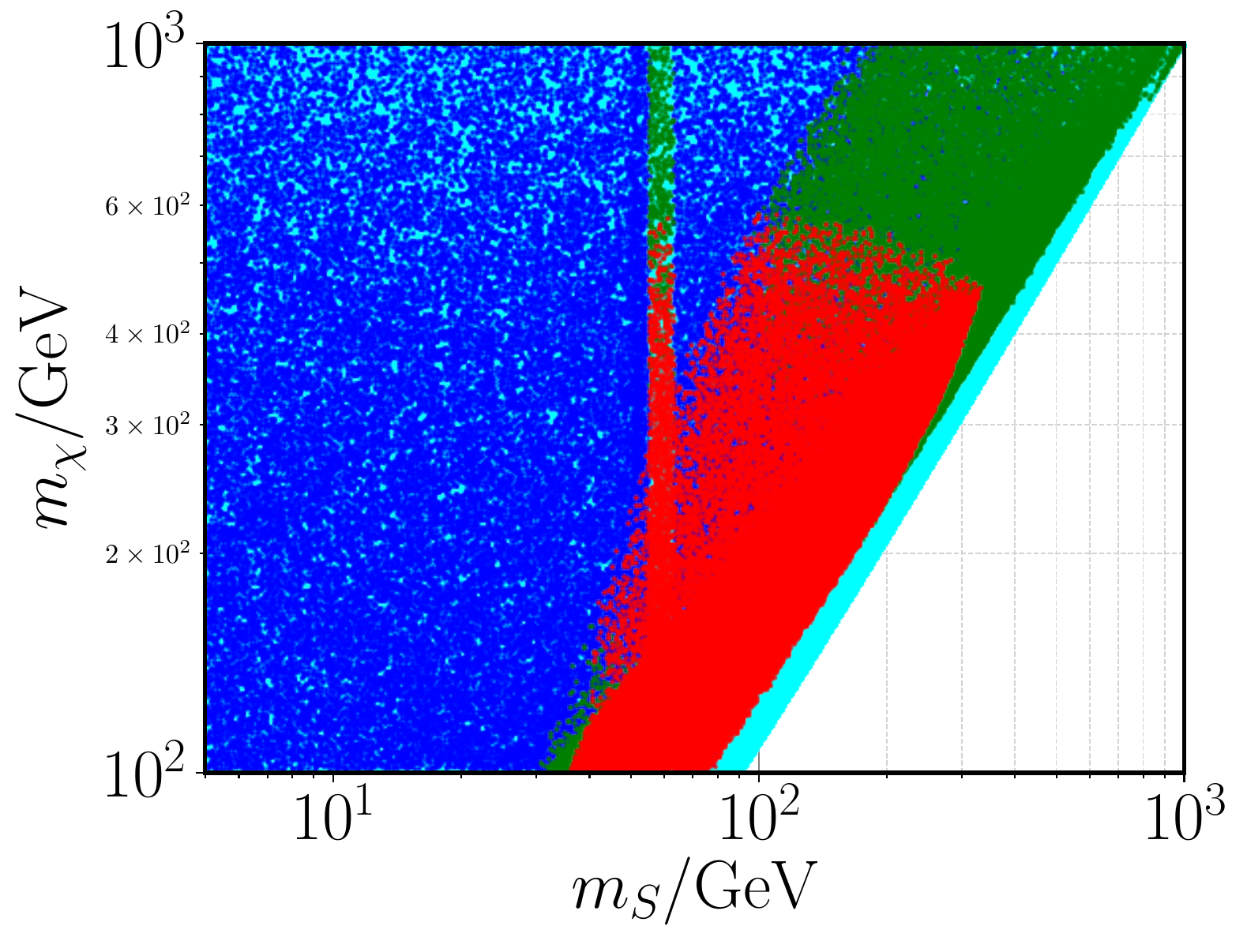}
	\includegraphics[scale=0.45]{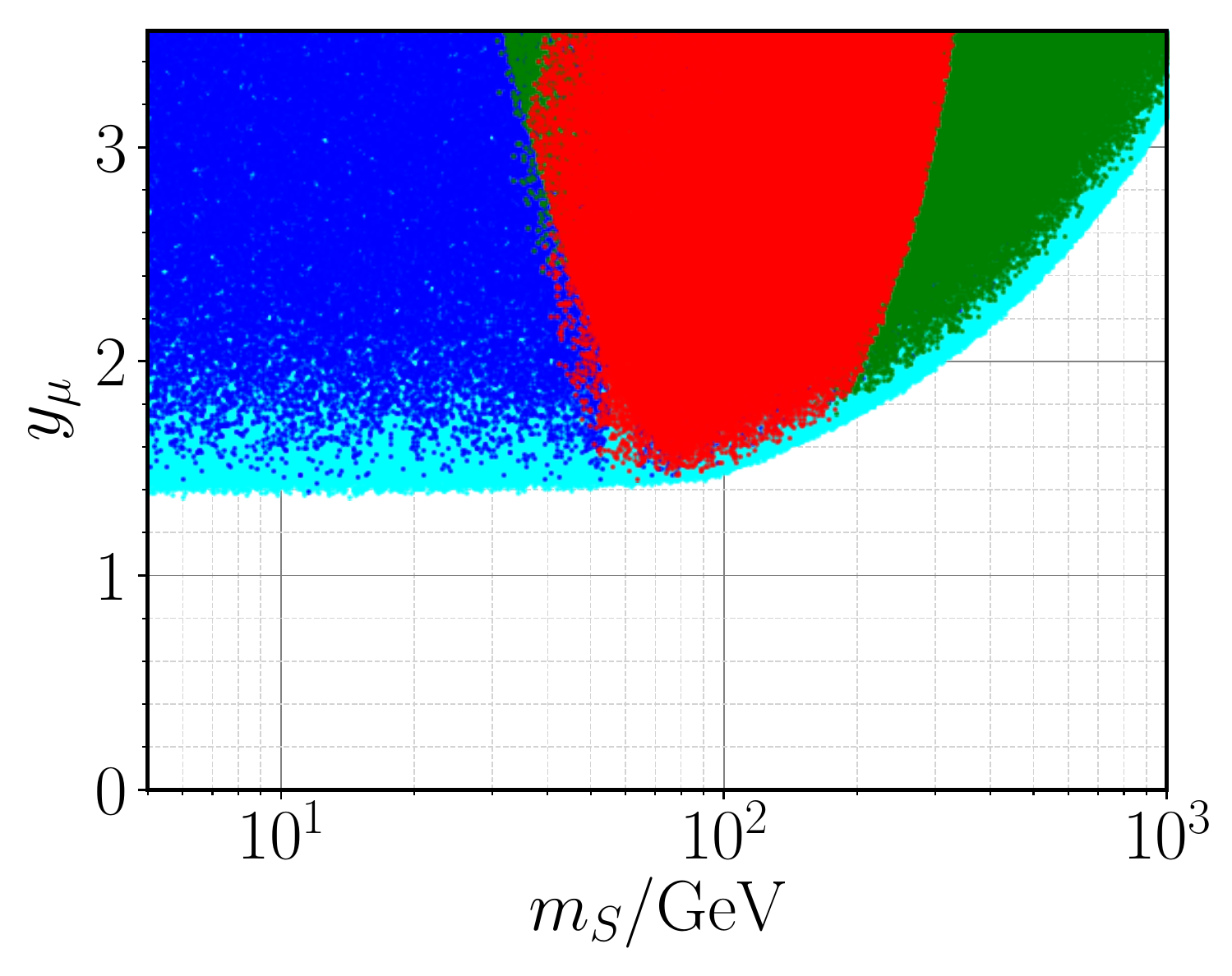}	
	\includegraphics[scale=0.45]{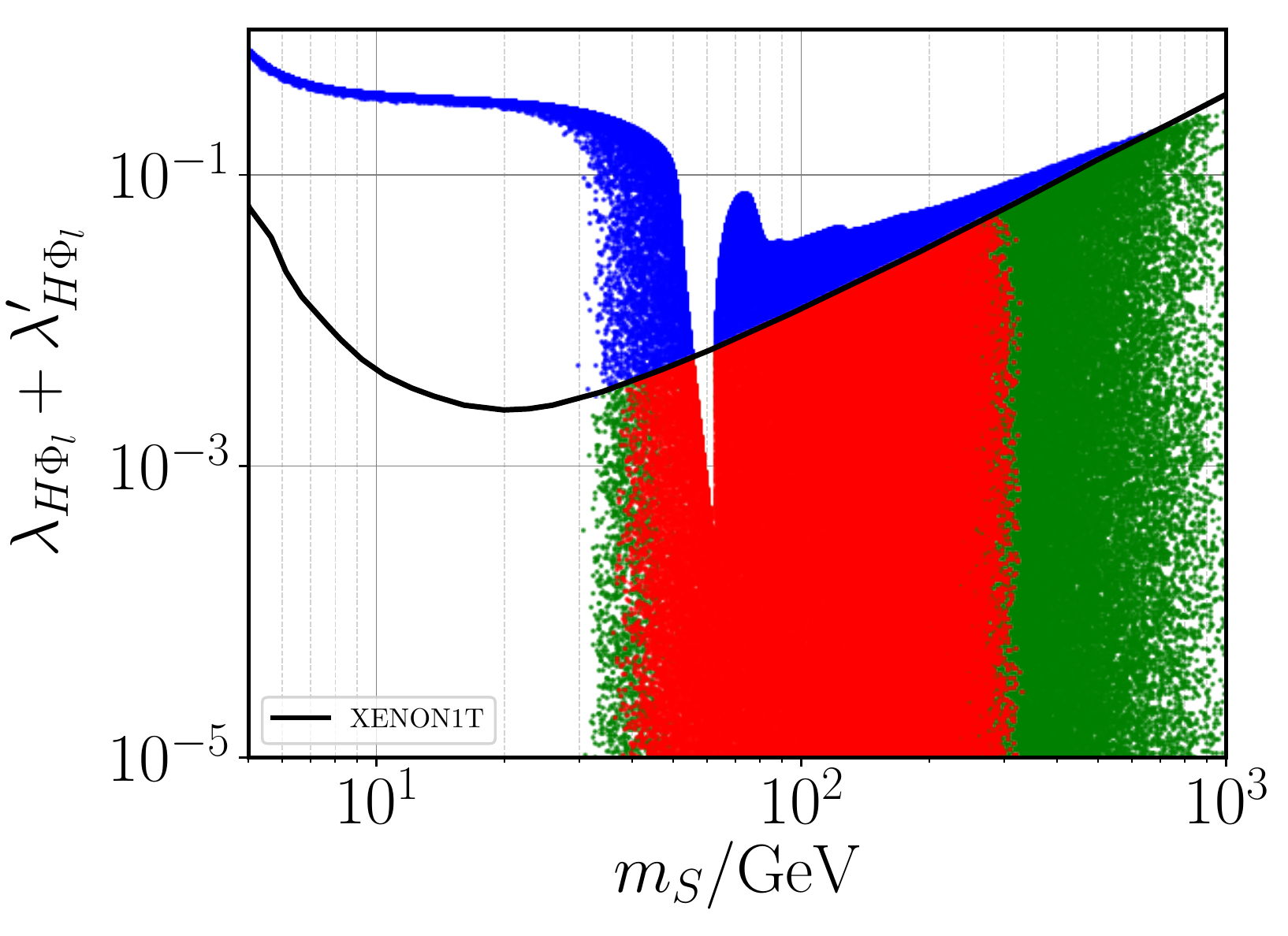}		
    \caption{Allowed parameter space projected in the planes  $m_S$-$m_A$ (Top Left), $m_S$-$m_\chi$ (Top Right), $m_S$-$y_\mu$ (Bottom Left) and $m_S$-$(\lambda_{H\Phi_l} + \lambda^\prime_{H\Phi_l})$ (Bottom Right). Other parameters are fixed as in Fig.~\ref{ScanFlavor}. }\label{ScanDM}
\end{figure}

In our numerical study, we perform a random scan of more than $10^9$ benchmark model points over the whole parameter space with the restrictions listed above. The final scanning results are shown in Figs.~\ref{ScanFlavor} and \ref{ScanDM}. First, all the colored points explain the $R(K^{(*)})$ associated anomalies while satisfying the ${\cal B}(B_s \to \mu^+ \mu^-)$ and $b\to s\gamma$ data within their $2\sigma$ confidence intervals. Second, when taking into account the observed DM relic abundance within $2\sigma$ CL range, the cyan colored points are excluded. The blue points correspond to those models which cannot satisfy the constraints from DM searches. In particular, the dominant experimental bound comes from the DM direct detection experiment XENON1T as evident from the lower right plot of Fig.~\ref{ScanDM}. Finally, the green points represent the models that are not allowed by the muon $(g-2)$ data within its $3\sigma$ range, while the red region is the common parameter space which can explain all the possible flavor and DM observations at the same time.

%%%%%%%%%%%%%%%%%%%%%%%%%

From Fig.~\ref{ScanFlavor}, it is clear that the $B$ meson decay data alone limits the dimensionless Yukawa coupling $|y_b|$ to be within the strip around 0.6, and $y_\mu$ to be greater than 1.38. The constraints from DM phenomenology, such as the DM relic density and direct detection searches, do not have a major impact on the parameter space, as can be seen by comparing the regions with cyan and blue points. Still, these DM constraints indeed limit the pseudoscalar meson mass to $m_A \gtrsim 50$~GeV. On the other hand, the inclusion of the muon $(g-2)$ data greatly reduces the allowed parameter space, with $0.25 \lesssim |y_b| \lesssim 0.65$, $1.4 \lesssim y_\mu \lesssim \sqrt{4\pi}$ and $m_\chi \lesssim 600$~GeV. This result is understandable, since the $3\sigma$ difference between the SM theoretical and experimental values of $\Delta a_\mu$ requires a mild suppression of the NP contribution in Eq.~(\ref{ExpG2}) forcing a not too large value of $m_\chi$.

%%%%%%%%%%%%%%%%%%%%%%%%%
%%%%%%%%%%%%%%%%%%%%%%%%%

In Fig.~\ref{ScanDM}, we show the same data points now in projections relevant to the DM physics. From these four plots, it is evident that the DM mass $m_S$ is confined to be in the range from about 30~GeV to 350~GeV, mainly due to the constraints from DM direct detections and to the muon $(g-2)$ anomaly. Moreover, the $m_S$-$m_\chi$ plot (the upper-right plot) shows an interesting feature: there are two distinct regions in the allowed parameter space (red points) corresponding to the two DM dominant freeze-out channels. The first one lies around the Higgs resonance $m_S \approx m_h/2$, meaning that the dominant DM annihilation at freeze-out is through the SM-like Higgs-mediated $s$-channel. Note that this channel is insensitive to the mass of $\chi$ that extends to 600~GeV which is the aforementioned muon $(g-2)$ limit on $m_\chi$. The other region starts at $m_\chi = 101.2$~GeV, where the DM mass lies in range $m_S \in [30~{\rm GeV}, 80~{\rm GeV}]$, and ends at $m_\chi \sim 600$~GeV corresponding to the DM mass range $m_S \in [140~{\rm GeV}, 350~{\rm GeV}]$. In this region there is a positive correlation between $m_S$ and $m_\chi$ dictated by Eq.~(\ref{SAnn1}) for 
the $t$- and $u$-channel process $SS\to \mu^+ \mu^-$. 
The bottom-right plot in the $m_S$-$(\lambda_{H\Phi_l}+\lambda_{H\Phi_l}^\prime)$ plane shows that the upper boundary of the colored region represents the largest value of the Higgs portal coupling $(\lambda_{H\Phi_l}+\lambda_{H\Phi_l}^\prime)$ with the correct DM relic abundance, corresponding to the cases with the Higgs-mediated process dominates over the DM annihilation during the freeze-out. The points below the boundary, beginning at around 30~GeV, are the ones where $\chi$-mediated process $SS\to \mu^+ \mu^-$ is the most important DM annihilation channel.
As a result, we find that except for the Higgs resonance region in which the upper boundary is allowed by the XENON1T data, the dominant DM annihilation channel for the DM generation is $SS\to \mu^+ \mu^-$. An important consequence of this result is that DM indirect detection searches~\cite{Feng:2010gw} are not expected to give any useful constraint to the present model, since the dominant DM annihilation cross section $\langle \sigma v\rangle_{SS\to \mu^+ \mu^-}$ is $d$-wave suppressed by its strong velocity dependence.

We finalize this section by noting that the choices $m_{\Phi_q} = 1.5$~TeV and chosen $y_s = -y_b/4$, when relaxed do not lead to any significant 
changes in the allowed parameter points that satisfy all constraints and explain all anomalies.

\section{Conclusion and Discussion}\label{Conclusion}
In the present work, we have explored a new class of particle physics solutions to lepton flavor universality violation observed in the decay $b\to s\mu^+ \mu^-$ by the LHCb and Belle Collaborations. At the same time we wanted a model with a good DM candidate and that would solve the muon $(g-2)$ anomaly. In order to achieve this goal, we have listed several simple extensions of the model in Ref.~\cite{Cerdeno:2019vpd} by restricting the new particle's $SU(2)_L$ representations to be either singlet, doublet or triplet and with $U(1)_Y$ hypercharges such that the electric charges
of the vectorlike fermion are 0 or $\pm 1$. For each model we have identified the possible DM candidates. We have thoroughly studied the flavor and DM phenomenology in one of the most promising models in this list, in which we introduce a $SU(2)_L$ doublet vector-like fermion $\chi$ and two complex scalar singlets, $\Phi_q$ and $\Phi_l$, the former is an $SU(3)_c$ triplet while the latter is colorless. 
As a result, the $R(K^{(*)})$ anomalies related to the $B$ meson decays can be solved by one-loop NP contributions, and the DM candidate can be the scalar component $S$ contained in $\Phi_l$.
By performing a random scan over the whole parameter space of physical interest, we have found that the combination of the XENON1T and $\Delta a_\mu$ data prefer a rather light DM candidate with its mass $m_S \in [30~{\rm GeV}, 350~{\rm GeV}]$. Moreover, the mass of the vectorlike fermionic mediator $\chi$ is restricted to be relatively light $m_\chi \lesssim 600$~GeV, and the Yukawa couplings should be sizeable with $y_\mu \gtrsim 1.4$ and $|y_b| \sim 0.6$.

\begin{figure}[!ht]
\centering
\includegraphics[width = 0.8 \linewidth]{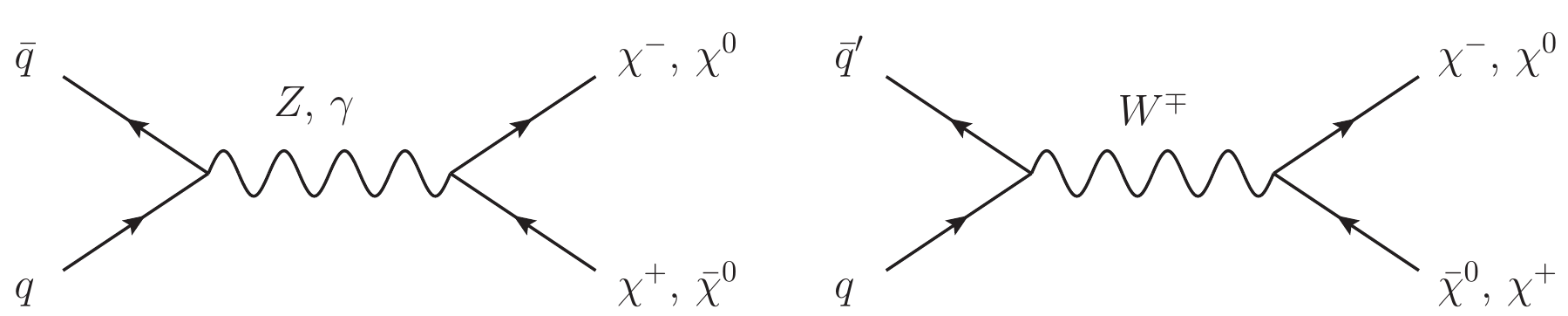}
\caption{Feynman diagrams for pair production of vectorlike fermions $\chi$. }\label{ChiP}
\end{figure}

Finally, we will briefly discuss possible collider searches of this model at the LHC. Since all new particles are $Z_2$-odd, ATLAS and CMS strategy should be to search for final states with leptons and jets plus DM particles, which are usually identified as the missing transverse energy $\slashed{E}_T$. One possible signal is vector-like lepton production mediated by $W^\pm$, $Z$ or $\gamma$, as illustrated in Fig.~\ref{ChiP}. 
The decay of $\chi^\pm$ ($\chi^0$) leads to the final states of $\mu^\pm S$ ($\nu_\mu S$)\footnote{If $A$ is lighter than $\chi$, the decays of $\chi$ into $\mu^\pm A$ and $\nu_\mu A$ are open, and $A$ can further decay through the three-body processes  $A\to \nu_\mu \bar{\nu}_\mu S$ and $A \to \mu^+ \mu^- S$, with the latter decay product observable at colliders. So here are additional LHC signatures, like $pp \to \chi^+ {\chi}^- \to 4\mu + \slashed{E}_T$ or $pp \to \chi^+ {\chi}^- \to 6\mu + \slashed{E}_T$. Also, the observed lepton spectra in the single-muon and dimuon channels would be modified due to the presence of $A$ decays. However, since $A$ is heavier than $S$ by assumption, the phase space of the decays $\chi \to A \mu/A\nu_{\mu}$ would be suppressed compared with $\chi \to S \mu/S\nu_{\mu}$. So it is expected that the latter decay channels dominate over the former ones.}. So we can consider the following LHC signals:
\begin{eqnarray}\label{SigChi}
pp &\to & \chi^+ \chi^- \to \mu^+ \mu^- + \slashed{E}_T\,,\nonumber\\
pp &\to & \chi^\pm \chi^0 \to \mu^\pm + \slashed{E}_T \,.
\end{eqnarray}
The cross sections for $\chi$ pair production at the 14~TeV LHC are shown in Table~\ref{ChiProd}.
\begin{table}[!th]\caption{Cross Sections for $\chi$ Pair Productions at 14~TeV LHC}
{\begin{tabular}{c|c|c}
~ & $m_\chi = 150$~GeV & $m_\chi = 500$~GeV \\
\hline
$\sigma(pp \to \chi^+ \chi^-)$/pb & 0.58 & $5.69\times 10^{-3}$ \\
$\sigma(pp \to \chi^- \bar{\chi}^0)$/pb & 0.67 & $5.36\times 10^{-3}$ \\
$\sigma(pp \to \chi^+ \chi^0)$/pb & 1.27 & $1.43\times 10^{-2}$ \\ 
\end{tabular}\label{ChiProd}}
\end{table}
From this table, if $\chi^{\pm}$ decays dominantly into $\mu^{\pm} S$, we can easily observe these two signals at the present LHC run, and even more at the future LHC High Luminosity run with its 3000~${\rm fb}^{-1}$ integrated luminosity.
Note that the above two signatures have been already investigated in the literature. In Ref.~\cite{Cerdeno:2019vpd} pair production of an $SU(2)_L$ doublet scalar with $Y=-1/2$ was studied. The Feynamn diagrams are almost the same as in Fig.~\ref{ChiP} with the fermion $\chi$ replaced by its scalar counterpart. Using ATLAS data~\cite{ATLAS:2017uun} and a leptonic Yukawa coupling equal to $2$, the DM candidate lighter than 30~GeV was excluded. Taking this Yukawa coupling to its perturbative limit $\sqrt{4\pi}$, the lower bound on the DM mass decreased to about 13~GeV. Since collider constraints on such channels are insensitive to the spin of the intermediated particles, we can apply these results to our case for reference.

\begin{figure}[!ht]
\centering
\includegraphics[width = 0.8 \linewidth]{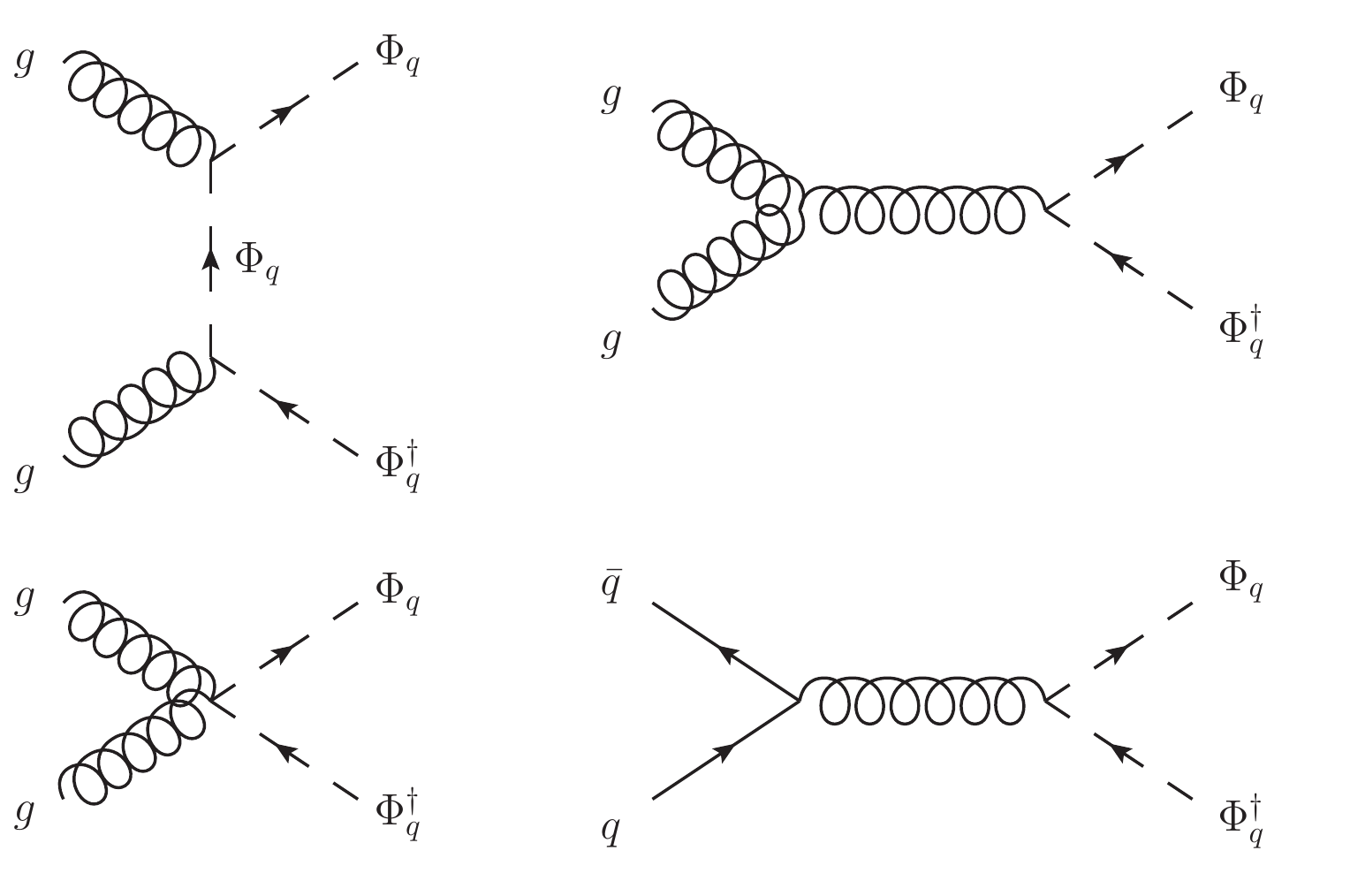}
\caption{Feynman diagrams for the pair production of colored scalar $\Phi_q$. }\label{PhiQ}
\end{figure}

Another interesting collider signature of this model is the pair production process of the colored $Z_2$-odd scalar $\Phi_q$. Note that $\Phi_q$ only couples to the second- and third-generation quarks by construction and that the dominant contribution to $\Phi_q \Phi_q^\dagger$ production at the LHC is though the pure QCD processes shown in Fig.~\ref{PhiQ}. The cross section has no dependence on the Yukawa couplings $y_b$ or $y_s$. 
Furthermore, since there is no tree-level coupling between $\Phi_q$ and the DM candidate $S$, $\Phi_q$ decays dominantly through the following cascade decay chains: $\Phi_q \to q\chi \to q S \mu (q S \nu_\mu) $ with the quark $q$ representing the second- and third-generation quarks. Therefore, possible signatures are
\begin{eqnarray}
pp \to \Phi_q \Phi_q^\dagger \to (jj + \mu^+ \mu^- + \slashed{E}_T)/ (jj +\mu^\pm + \slashed{E}_T)/ (jj + \slashed{E}_T)\,,
\end{eqnarray}
where $j$ denotes jets in the final states. A simple numerical study of the $\Phi_q \Phi^\dagger_q$ production at the LHC gives its cross section to be $\sigma(pp \to \Phi_q \Phi_q^\dagger) = 1.33\times 10^{-4}$~pb for $m_{\Phi_q} = 1.5$~TeV. By taking into account the fact that nearly half of $\Phi_q$ goes to the final state $t(c)\bar{\chi}^0$ and the other half to $b(s) \chi^+$, it is still possible to observe the signals above at the HL run. For the ${\rm dijet} + \slashed{E}_T$ search, a similar scenario was carefully discussed in Ref.~\cite{Cerdeno:2019vpd} by using recent LHC data~\cite{Aaboud:2017rzf}, with the minor difference that the colored scalar was a $SU(2)_Y$ doublet with $Y=1/6$ there. The general conclusion was that, for a light DM particle, the colored scalar $\Phi_q$ with $m_{\Phi_q} \lesssim 1~{\rm TeV}$  was excluded by the current LHC data. This result can be directly applied to our case here since the main production mechanism of the exotic colored states is the same. On the other hand, our present model predicts that the final states of $(jj + \mu^+ \mu^- + \slashed{E}_T)$ and $(jj +\mu^\pm + \slashed{E}_T)$ should have almost equal cross sections as $jj + \slashed{E}_T$, but, due to the presence of additional muons, these two channels are more promising to be measured and probed at the LHC.
% However, a detailed study of the collider searches for the present model is beyond the scope of this work, so we would like to explore this aspect in the near future.

%%%%%%%%%%%%%%%%%%%%%%%%%%%%%%%%%%%%%%%
%%%%%%%%%%%%%%%%%%%%%%%%%%%%%%%%%%%%%%%%%%%%%%%%%%%%%%%%%%%%%%%%%%%%%%%%%%%%%%%%%%%%%%%%%%%%%%%%%%%%%%%%%%%%%%%%%%%%%%%%%%%%%%%%%%%
\section*{Acknowledgments}
DH and APM are supported by the Center for Research and Development in Mathematics and Applications (CIDMA) through the Portuguese Foundation for Science and Technology (FCT - Fundação para a Ciência e a Tecnologia), references UIDB/04106/2020 and UIDP/04106/2020. DH, APM and RS are supported by the project PTDC/FIS-PAR/31000/2017. DH is also supported by the Chinese Academy of Sciences (CAS) Hundred-Talent Program. RS is also supported by FCT, Contracts UIDB/00618/2020, UIDP/00618/2020, and CERN/FISPAR/0002/2017, and by the HARMONIA project, contract UMO-2015/18/M/ST2/0518. APM is also supported by the project CERN/FIS-PAR/0027/2019 and by national funds (OE), through FCT, I.P., in the scope of the framework contract foreseen in the numbers 4, 5 and 6 of the article 23, of the Decree-Law 57/2016, of August 29, changed by Law 57/2017, of July 19.

%\section*{References}

\end{document}